\documentclass[10pt,journal,compsoc]{IEEEtran}

\usepackage{url}
\usepackage{graphicx}
\usepackage{color}
\usepackage{epsfig}
\usepackage{latexsym}

\usepackage{amssymb}
\usepackage{amsmath}
\usepackage{amsthm}
\usepackage{framed}
\usepackage{enumerate}
\usepackage{makeidx}
\usepackage{bm}
\usepackage{epsfig}
\usepackage{graphics}
\usepackage{colortbl}
\usepackage{colordvi}
\usepackage{verbatim}
\usepackage{amsmath}
\usepackage{multirow} 
\usepackage{booktabs} 

\usepackage{subfig}

\newcommand{\probconverge}{\mbox{~$\stackrel{P}{\longrightarrow}$}}

\newcommand{\bT}{\mathbb{T}}
\newcommand {\C} {{\rm I\kern-5.5pt C}}

\newcommand{\bP}[1]{{\mathbb{P}}\left[{#1}\right]}
\newcommand{\bE}[1]{{\mathbb{E}}\left[{#1}\right]}

\newcommand{\1}[1]{{\bf 1}\left[#1\right]}       % indicator 1{...}

\newcommand{\fsquare}{\vrule height6pt width7pt depth1pt}   % filled square
\newcommand{\myproof}{{\hfill \\ \bf Proof. \ }}           % Proof
\newcommand{\myendpf}{\hfill\fsquare \\[0.1in]}             % end of proof
\newcommand{\myvec}[1]{{\mbox{\boldmath{$#1$}}}}

\newtheorem{theorem}{Theorem}[section]

\newtheorem{lemma}[theorem]{Lemma}
\newtheorem{proposition}[theorem]{Proposition}

\newtheorem{fact}[theorem]{Fact}
\newtheorem{assumption}{Assumption}

\allowdisplaybreaks

\begin{document}

%\sloppy

\title{Asymptotic degree distributions \\
         in large (homogeneous) random networks:\\
        A little theory and a counterexample
        \thanks{This work was supported in part by NSF Grant CCF-1217997.
The paper was completed during the academic year 2014-2015 while A.M. Makowski 
was a Visiting Professor with the Department of Statistics of the Hebrew University of Jerusalem 
with the support of a fellowship from the Lady Davis Trust.
% The authors also
%thank a colleague (who wishes to remain anonymous) for suggestions that lead to a
%shorter proof of the one law in Theorem \ref{thm:OneLaw}.
}
\thanks{This document does not contain technology or technical data controlled 
under either the U.S. International Traffic in Arms Regulations or the U.S. Export Administration Regulations. }
\thanks{
Parts of the material were presented
in the 53rd IEEE Conference on Decision and Control (CDC 2015),
Osaka (Japan), December 2015.
}
}

\author{Siddharth Pal
\thanks{S. Pal was with the Department of Electrical and Computer Engineering, and 
the Institute for Systems Research, University of Maryland, College
Park, MD 20742 USA. He is now with Raytheon BBN Technologies (email: siddharth.pal@raytheon.com). }\
and Armand M. Makowski
\thanks{A. M. Makowski is with the
Department of Electrical and Computer Engineering, and the
Institute for Systems Research, University of Maryland, College
Park, MD 20742 USA (e-mail: armand@isr.umd.edu).}
}

\IEEEtitleabstractindextext{
\begin{abstract}
%\normalsize 

%In random graph models, the degree distribution of an {\em individual} node should be distinguished from
%the (empirical) degree distribution of the {\em graph} that records the fractions of nodes with given degree.
%We introduce a general framework to explore  when these two  degree distributions coincide asymptotically in a sequence
%of {\em homogeneous} random networks of increasingly large size.  
%The discussion is carried under three basic statistical assumptions on the degree sequences: 
%(i) a weak form of  distributional homogeneity;
%(ii) the existence of an asymptotic (nodal) degree distribution; 
%and (iii) a weak form of asymptotic uncorrelatedness.
%It follows from the discussion that under (i)-(ii) the asymptotic equality of the two degree distributions occurs if and only if (iii) holds.
%We use this observation to show that the asymptotic equality may fail in some homogeneous random networks.
%The counterexample is found  in the class of random threshold graphs with exponentially distributed fitness 
%for which (i) and (ii) hold but where (iii) does not.
%An implication of this finding is that these random threshold graphs  cannot be used as a substitute to the Barab\'asi-Albert model
%for scale-free network modeling, as was proposed by some authors.
%The results can also be formulated for non-homogeneous models by making use of  a random sampling procedure
%over the nodes. This approach is now applicable to many classes of network models, including preferential attachment models and 
%locally weak convergent sequence of random graphs.

In random graph models, the degree distribution of an {\em individual} node should be distinguished from the (empirical) degree distribution of the {\em graph} that records the fractions of nodes with given degree. We introduce a general framework to explore  when these two  degree distributions coincide asymptotically in a sequence
of {\em homogeneous} random networks of increasingly large size.  
The discussion is carried under three basic statistical assumptions on the degree sequences: (i) distributional homogeneity; (ii) existence of an asymptotic (nodal) degree distribution; and (iii) asymptotic uncorrelatedness. It follows from the discussion that under (i)-(ii) the asymptotic equality of the two degree distributions occurs if and only if (iii) holds. We use this observation to show that the asymptotic equality may fail in some homogeneous random networks. The counterexample is found in the class of random threshold graphs for which (i) and (ii) hold but where (iii) does not. An implication of this finding is that these random threshold graphs  cannot be used as a substitute to the Barab\'asi-Albert model for scale-free network modeling, as was proposed by some authors. The results can also be formulated for non-homogeneous models by making use of  a random sampling procedure over the nodes.

\end{abstract}

}

\maketitle
\IEEEdisplaynontitleabstractindextext
\IEEEpeerreviewmaketitle

\IEEEraisesectionheading{\section{Introduction}
\label{sec:Introduction}}

In the past three decades considerable efforts have been devoted to understanding 
the rich structure and functions of complex networks, be they technologically engineered,
found in nature or generated through social interactions.
These developments have been recorded in surveys,
e.g., \cite{AlbertBarabasiReview, DorogovstevMendes, NewmanSurvey},
research monographs, e.g.,
\cite{BarratBarthelemyVespignani, CohenHavlinBook, Durrett_Book, JacksonBook, NewmanBook},
and anthologies of research papers, e.g., \cite{NewmanBarabasiWatts}.

The questions of  interest often relate to a collection of entities (alternatively called nodes, 
agents, etc.) and to a set of relationships between them. 
The pairings can be physical, logical or social in nature; when pictured 
as links or edges between nodes, they naturally give rise to {\em graphs} and 
graph-like structures (customarily referred to as networks) on the set of  nodes. Often the pairwise 
relationships are best viewed as inherently random, suggesting that 
random graph models be used to frame the relevant issues -- Here we understand a random graph to be a 
graph-valued random variable (rv).

A popular research direction has been concerned with designing 
random graph models that exhibit key properties observed in real networks.
Historically attention has been given
to the simplest of network properties, namely the degree of nodes and their various distributions.
The discussion invariably starts with the work 
of Erd\H{o}s and R\'enyi \cite{ER1960}: With $n$ nodes and link probability $p$,
the (binomial) Erd{\H{o}}s-R{\'e}nyi graph $\mathbb{G}(n;p)$ postulates that
the $\frac{n(n-1)}{2}$ potential undirected links between these $n$ nodes
are each created with probability $p$, independently of each other. 
The degree distribution in Erd{\H{o}}s-R{\'e}nyi graphs is announced to be Poisson-like, the justification going roughly as follows:
(i) With $D_{n,k} (p)$ denoting the degree rv of node $k$ in $\mathbb{G}(n;p)$, 
the rvs $D_{n,1} (p), \ldots , D_{n,n} (p)$ 
are identically distributed,  each distributed according to
a binomial rv ${\rm Bin}(n-1;p)$;
(ii) If the link probability scales with $n$ as $p_n \sim \frac{\lambda}{n}$ for some $\lambda > 0$, then
Poisson convergence ensures the distributional
convergence
\begin{equation}
D_{n,1}(p_n) \Longrightarrow_n D
\label{eq:PoissonConvergenceER}
\end{equation}
with $D$ denoting a Poisson rv with parameter $\lambda$.
A rich asymptotic theory has been developed
for Erd\H{o}s-R\'enyi graphs in the many node regime; see the monographs
\cite{Bollobas_Book, DraiefMassoulieLMMS, Durrett_Book, JansonLuczakRucinski}.

However, in many networks the data tells a different story: If the network
comprises a large number $n$ nodes and $N_n(d)$ is the number of nodes with degree $d$ in the network, 
then statistical analysis suggests a power-law behavior of the form
\begin{equation}
\frac{ N_n(d) }{n} \simeq C d^{-\alpha}
\label{eq:PowerLaw}
\end{equation}
for some $\alpha$ in the range $[2,3]$ (with occasional exceptions) and $C>0$.
See \cite[Section 4.2]{Durrett_Book} for an introductory discussion and references, 
and the paper by Clauset et al. \cite{ClausetShaliziNewman} for a principled statistical framework.
Statements such as (\ref{eq:PowerLaw})
are usually left somewhat vague as the range of $d$ is never carefully specified;
networks where (\ref{eq:PowerLaw}) was observed are often called {\em scale-free} networks.

On account of this observation, Erd{\H{o}}s-R{\'e}nyi graphs were deemed inadequate for
modeling scale-free networks (as well as other networks of interest).
As a result, new classes of random graph models have been  proposed in an attempt to capture the
behavior (\ref{eq:PowerLaw}) (and other properties), 
e.g., the configuration model \cite{BenderCaulfield_1978, Bollobas_1980, MolloyReed_1995, MolloyReed_1998}, 
generalized random graphs \cite{BrittonDeijfenMartinLof_2006}, and exponential random graphs 
\cite{HollandLeinhardt_1981, Strauss_1986} to name some of the possibilities.
The Barab\'asi-Albert network model came to prominence 
for its ability to formally \lq\lq explain" the existence of power law degree distributions in large networks
via the mechanism of preferential attachment \cite{BarabasiAlbert}.

The statement (\ref{eq:PowerLaw})
concerns an {\em empirical}  degree distribution computed {\em network-wide}, whereas the convergence
(\ref{eq:PoissonConvergenceER}) addresses the behavior of the (generic) degree of a {\em single} node, 
its distribution being identical across nodes.
A natural question is then whether these two different points of view are compatible with each other and 
can be reconciled, at least asymptotically, in large networks, and if so, under what conditions.
The purpose of this paper is to explore this issue in some details.
What follows is an outline of some of the contributions along these lines:

\noindent
{\bf 1.} In Section \ref{sec:Framework} a  general framework to investigate this discrepancy
          is introduced  in terms of a sequence of random graphs $\{ \mathbb{G}_n, \ n=1,2, \ldots \}$
          whose size goes to infinity with $n$. Two different settings of increasing generality are considered.
 
\noindent
{\bf 2.} The {\em homogeneous} setting
          captures situations where an asymptotic nodal degree distribution exists, and is presented  in Section \ref{sec:HomogeneousAssumptions}.
          It is defined in terms of the following three assumptions:
          \begin{enumerate} 
          \item[(i)] A weak form of distributional homogeneity (hence the terminology homogeneous networks):  
           In particular, for each $n=1,2, \ldots $, the degree rvs in $\mathbb{G}_n$ 
           are identically distributed across nodes -- Let $D_n$ denote 
           the generic degree rv in $\mathbb{G}_n$;
           \item[(ii)] Existence of an asymptotic (nodal) degree distribution:  In analogy with (\ref{eq:PoissonConvergenceER}),
           there exists an $\mathbb{N}$-valued rv $D$ such that
           \begin{equation}
           D_n \Longrightarrow_n D.
           \label{eq:ConvergenceGeneric}
           \end{equation}
           Let $( p(d), \ d=0,1, \ldots )$ denote the pmf of $D$; and
           \item[(iii)] Asymptotic uncorrelatedness: The degree rvs $\{ D_n, \ n=1,2, \ldots \}$ 
           display a weak form of asymptotic \lq\lq pairwise independence."
           \end{enumerate}

\noindent
{\bf 3.}  The relevant  results for the homogeneous case are discussed in  Section \ref{sec:HomogeneousLittleTheory}. 
              Under the aforementioned assumptions, Proposition \ref{prop:ConvergenceEmpirical+PMF} states that 
              if $( P_n(d), \ d=0,1, \ldots )$ is the empirical degree distribution 
              in $\mathbb{G}_n$ (with $P_n(d)$ denoting the fraction of nodes with degree $d$ in $\mathbb{G}_n$), then
              \begin{equation}
             P_n (d) \probconverge_n ~ p(d), \quad d=0,1, \ldots
             \label{eq:AsymptoticEmpiricalPMF}
             \end{equation}
             where the pmf $( p(d), \ d=0, 1, \ldots )$ on $\mathbb{N}$ is as postulated in (ii) above. 
             A strengthening of this result in terms of total variation distance
             is provided as Proposition \ref{prop:ConvergenceEmpirical+TotalVariation}.

\noindent
{\bf 4.}  A more general setting is considered in Section \ref{sec:GeneralSettingLittleTheory} where degree homogeneity, namely (i) above,
             is replaced by a {\em random sampling} procedure over pairs of nodes.
             Many situations are easily fitted into this more general framework. They include
             the {\em non}-homogeneous Barab\'asi-Albert model (and other growth models), sequences of
             deterministic graphs and sequences which are {\em locally weakly convergent} \cite{AldousSteele}
            (or weakly convergent in the sense of Benjamini and Schramm \cite{BenjaminiSchramm}).

\noindent
{\bf 5.}  In Section \ref{sec:CommonSetting} we introduce a broad class of models where the underlying assumptions (i)--(iii) can be checked;
             this provides a natural and convenient setting for applying Proposition \ref{prop:ConvergenceEmpirical+PMF}.
             Erd{\H{o}}s-R{\'e}nyi graphs (under the scaling yielding (\ref{eq:PoissonConvergenceER})) are readily subsumed in this framework, 
            as are many other homogeneous networks of interest in applications; see \cite{SPal_Thesis} for details.
            This resolves the discrepancy mentioned earlier in that the appropriate version of
            (\ref{eq:AsymptoticEmpiricalPMF}) does hold for both Erd{\H{o}}s-R{\'e}nyi graphs 
            (by virtue of Proposition \ref{prop:ConvergenceEmpirical+PMF})
            \cite{SPal_Thesis} and for the Barab\'asi-Albert model (for which (\ref{eq:AsymptoticEmpiricalPMF}) holds
            with limiting pmf satisfying $p(d) \sim d^{-3}$ ($d \rightarrow \infty)$ \cite{Bollobas_BAmodel}).

\noindent
{\bf 6.}   Next we turn our attention to the proposition, too often taken for granted,
          that in homogeneous random graphs the convergence (\ref{eq:ConvergenceGeneric}) of the generic degree distribution 
           {\em automatically} implies the convergence (\ref{eq:AsymptoticEmpiricalPMF}) of the empirical degree distribution.  
          In Section \ref{sec:CounterExample} we provide a counterexample drawn from 
          the class of random threshold graph models \cite{CCDM, FIKMMU, MakowskiYagan-JSAC, SC}.
          For this class of models under {\em exponentially} distributed fitness, 
          although  (\ref{eq:ConvergenceGeneric}) is known to take place with $p(d) \sim d^{-2}$ ($d \rightarrow \infty)$ \cite{FIKMMU},
          we show that (\ref{eq:AsymptoticEmpiricalPMF}) fails to hold. 
          This fact, contained in Proposition \ref{prop:NonConvergenceInProbability}, constitutes an easy byproduct of Proposition \ref{prop:AssumptionC+fails}.
          Proofs occupy Section \ref{sec:ProofPart+1} to Section \ref{sec:ProofsAsymptoticsOrderStatistics},
          and rely on the asymptotics of
          order statistics for i.i.d. variates \cite{EKM, LeadbetterLindgrenRootzen}.
          We illustrate this failure through limited simulation results in Section \ref{sec:SimulationResults}.

\noindent
{\bf 7.}   One implication of this last finding is that random threshold graphs with exponentially distributed fitness
         {\em cannot} be used as an alternative scale-free model to the Barab\'asi-Albert model (see below) as claimed by some authors
         \cite{CCDM,SC}. Indeed, only the convergence (\ref{eq:AsymptoticEmpiricalPMF}) has meaning 
         in the preferential attachment model
         while  (\ref{eq:ConvergenceGeneric}) is meaningless there, with the situation being reversed for random threshold graphs. In other words,
         leaving aside the issue of which value of $\alpha$ is appropriate,
         the two models cannot be compared in terms of their degree distributions!
         This highlights the fact that {\em even} in homogeneous graphs, 
         the network-wide degree distribution and the nodal degree distribution may capture vastly different information.  

Some of the results discussed in this paper were announced in the conference paper \cite{PalMakowski_CDC2015}, mostly
without proofs. Different proofs to establish Proposition \ref{prop:NonConvergenceInProbability}.
were originally given in the Ph.D. thesis of the first author \cite{SPal_Thesis}.

\section{A simple framework}
\label{sec:Framework}

First some notation and conventions:
The random variables (rvs) under consideration are all 
defined on the same probability triple  $(\Omega, {\cal F}, \mathbb{P})$. 
The construction of a probability triple sufficiently large to carry all the required rvs is standard,
and omitted in the interest of brevity.
All probabilistic statements are made with respect to the probability
measure $\mathbb{P}$, and we denote the corresponding expectation
operator by $\mathbb{E}$. 
The notation $\probconverge_n$ (resp.  $\Longrightarrow_n$) 
is used to signify convergence in probability
(resp. convergence in distribution) (under $\mathbb{P}$) with $n$ going to infinity;
see the monographs \cite{BillingsleyBook, ChungBook} for definitions and properties.
If $E$ is a subset of $\Omega$, then $\1{E}$ is the indicator rv
of the set $E$ with the usual understanding that $\1{E}(\omega) = 1$ (resp. $\1{E}(\omega) = 0$) if $\omega \in E$
(resp. $\omega \notin E$).
%With $X$ and $Y$ two $\mathbb{R}$-valued rvs, we say that  $X$ is smaller in the strong stochastic order than $Y$,
%written $X \leq_{st} Y $,  if $\bP{ X > t } \leq \bP{ Y > t }$ for all $t$ in $\mathbb{R}$. 
The symbol $\mathbb{N}$ (resp. $\mathbb{N}_0$)  denotes the set
of non-negative (resp. positive) integers.

The discussion is carried out in the following framework often encountered in the literature; see Section \ref{sec:CommonSetting} for examples:
Given is a sequence of random graphs $\{ \mathbb{G}_n , \ n=2,3, \ldots \}$ defined 
on the probability triple $(\Omega, {\cal F}, \mathbb{P})$ -- We interchangeably use the terms random graphs and random networks.
Fix $n=2,3, \ldots$. The random graph $\mathbb{G}_n$ is then an ordered pair
$(V_n, \mathbb{E}_n)$ defined on the set of nodes $V_n $ with random edge set $\mathbb{E}_n \subseteq V_n \times V_n$.
Throughout the deterministic set $V_n$ is assumed to be non-empty and finite.
The random edge set $\mathbb{E}_n$ is equivalently determined by 
a set of $\{0,1\}$-valued edge rvs $\{ \chi_n (k,\ell), \ k,\ell \in V_n \}$ -- Thus,
$\chi_n(k,\ell) =1$ (resp. $\chi_n(k,\ell) =0$) if there is a directed edge (resp. no edge) {\em from} node $k$ {\em to} node $\ell$,
so that
$ \mathbb{E}_n = \{ (k,\ell) \in V_n \times V_n: \ \chi_n (k,\ell) = 1 \}$. 
We do not necessarily assume that $\mathbb{G}_n$ is an {\em undirected} graph, and we allow {\em self-loops}. 
There is no loss in generality in taking $V_n = \{ 1, \ldots , k_n \}$ for some positive integer $k_n$.
In most cases of interest  $V_n = \{ 1, \ldots , n \}$ so that $k_n = n $.

For each $k$ in $V_n$, the degree of node $k$ in the random graph $\mathbb{G}_n$ is the rv $D_{n,k}$ given by
\begin{equation}
D_{n,k} = \sum_{\ell \in V_n} \chi_n (k,\ell).
\label{eq:Out-degreeDefinition}
\end{equation}
For each $d=0,1, \ldots $, the rv $N_n(d)$ defined by
\begin{equation}
N_n(d) = \sum_{k\in V_n} \1{ D_{n,k}= d }
\label{eq:NumberNodes}
\end{equation}
counts the number of nodes in $V_n$ which have degree $d$
in $\mathbb{G}_n$. The fraction of nodes in $V_n$ with degree $d$ in
$\mathbb{G}_n$ is then given by
\[
P_n(d) = \frac{ N_n(d) }{ |V_n| } = \frac{1}{|V_n|} \sum_{k\in V_n} \1{ D_{n,k}= d }.
\]
This defines the random pmf
\[
\myvec{P}_n = \left ( \frac{ N_n(d) }{ |V_n| }, \ d=0,1, \ldots \right )
\]
on $\mathbb{N}$ with support contained in $V_n \cup \{ 0 \}$.
Strictly speaking, the expression (\ref{eq:Out-degreeDefinition}) defines the {\em out}-degree of a node.
However, everything said for out-degrees can also be developed for in-degrees with no substantive changes.
In what follows the term degree will refer interchangeably to either out-degree or in-degree,
the point being moot when considering undirected graphs as is the case in many situations.

For each $d=0,1, \ldots $, we explore the convergence (in probability) of the random sequence  $\{ P_n(d), \ n=2, \ldots \}$ 
to a deterministic limit, say $L(d)$ in $\mathbb{R}$, when the graph size becomes infinitely large, namely 
$\lim_{n \rightarrow \infty} |V_n| = \infty$. 
%Equivalently, we assume  the sequence  $n \rightarrow k_n$ to be monotone increasing with $\lim_{n \rightarrow \infty} k_n = \infty$. 

For sequences of bounded rvs, convergence in probability and mean-square convergence  are equivalent by standard
facts concerning modes of convergence for rvs \cite{BillingsleyBook, ChungBook}.
Therefore, the convergence
 \begin{equation}
P_n(d) \probconverge_n ~ L(d)
\label{eq:LIMIT}
\end{equation}
occurs if and only if
\begin{equation}
\lim_{n \rightarrow \infty}  \bE{ \left | P_n(d)  - L(d) \right |^2 } = 0.
\label{eq:MSConvergence}
\end{equation}
For each $n=2,3, \ldots $, standard properties of the variance give
\begin{eqnarray}
\lefteqn{ \bE{ \left | P_n(d)  - L(d) \right |^2 }  } & &
\nonumber \\
&=&
{\rm Var} \left [ P_n(d) \right ] 
+
 \left | \bE{ P_n(d) }   - L(d) \right |^2 ,
\label{eq:VARIANCE}
\end{eqnarray}
and the following characterization is readily obtained.

\begin{fact}
{\sl With $d=0,1, \ldots$, the convergence in probability (\ref{eq:LIMIT}) occurs
to some scalar $L(d)$ if and only if we simultaneously have
$\lim_{n \rightarrow \infty} \bE{P_n(d) } = L(d)$
and $\lim_{n \rightarrow \infty} {\rm Var} \left [ P_n(d) \right ] = 0$.
}
\label{fact:BasicFact}
\end{fact}

To exploit this observation we begin by computing the first two moments $\bE{ N_n(d) }$ and $\bE{ N_n(d)^2 } $.
The definition (\ref{eq:NumberNodes}) of the rv $N_n(d)$ yields the expressions
\begin{equation}
\bE{ N_n(d) }
=  \sum_{k \in V_n} \bP{ D_{n,k} = d } 
\label{eq:MEAN}
\end{equation}
and \begin{eqnarray}
\lefteqn{ \bE{ N_n(d)^2 } } & &
\nonumber \\
&=&
\sum_{k \in V_n} \bP{ D_{n,k} = d } 
\label{eq:SECOND} \\
& &
+~ \sum_{k \in V_n } \left ( \sum_{\ell \in V_n : \ \ell \neq k} \bE{  \1{ D_{n,k} = d }   \1{ D_{n,\ell} = d }  } \right )
\nonumber
\end{eqnarray}
by the binary nature of the involved rvs.

We leverage Fact \ref{fact:BasicFact} in two different settings:
The {\em homogeneous} setting, introduced in  Section \ref{sec:HomogeneousAssumptions},
captures situations already mentioned in the introduction where an asymptotic nodal degree distribution exists; the relevant  results are
presented in  Section \ref{sec:HomogeneousLittleTheory}.
A more general setting is considered in Section \ref{sec:GeneralSettingLittleTheory}.

\section{The homogeneous case -- Assumptions}
\label{sec:HomogeneousAssumptions}

First we specify what we mean by a random network (or interchangeably, 
a random graph) to be {\em homogeneous} for the purpose of this paper.

\begin{assumption}
{\sl (Homogeneity) For each $n=2,3, \ldots$, the degree rvs
in $\mathbb{G}_n$ are equidistributed in the sense that
\begin{equation}
D_{n,k} =_{st} D_{n,1},
\quad k \in V_n
\label{eq:GenericDegreeRV}
\end{equation}
and
\begin{equation}
(D_{n,k}, D_{n,\ell} )  =_{st} ( D_{n,1},D_{n,2} )
\quad
\begin{array}{c}
k \neq \ell \\
k,\ell  \in V_n .\\
 \end{array}
 \label{eq:GenericDegreeRV_Pair}
\end{equation}
}
\label{ass:A}
\end{assumption}

Obviously condition (\ref{eq:GenericDegreeRV_Pair}) implies condition (\ref{eq:GenericDegreeRV}).
In many settings (see Section \ref{sec:CommonSetting}),
Assumption \ref{ass:A} follows from the stronger structural assumption that
for each $n=1,2, \ldots $, the edge rvs $\{ \chi_n (k,\ell) , \ k, \ell  \in V_n \}$ (or a subset thereof in the undirected case)
are {\em exchangeable} -- Random networks with this property are traditionally called homogeneous.
Under Assumption \ref{ass:A}, for each $n=2,3, \ldots $, 
it is appropriate to speak of {\em the} degree distribution of a node in $\mathbb{G}_n$,
namely the distribution of $D_{n,1}$.

In many cases of interest the degree rvs $\{ D_{n,1}, \ n=2,3, \ldots \}$ converge in the following sense.

\begin{assumption}
{\sl (Existence of an asymptotic degree distribution) 
Assume that Assumption \ref{ass:A} holds and that there exists an $\mathbb{N}$-valued rv $D$ such that
\begin{equation}
D_{n,1} \Longrightarrow_n D.
\label{eq:ConvergenceToLittleP}
\end{equation}
Let $ \myvec{p} = ( p(d), \ d =0,1, \ldots )$ denote the pmf of the limiting rv $D$.
}
\label{ass:B}
\end{assumption}

Assumption \ref{ass:B} can be rephrased as 
\begin{equation}
\lim_{n \rightarrow \infty} \bP{ D_{n,1} = d } = p(d),
\quad 
d=0,1, \ldots 
\label{eq:ConvergenceDegreeDistribution}
\end{equation}
Even in well-structured settings where Assumption \ref{ass:A} holds,
the convergence (\ref{eq:ConvergenceToLittleP}) may fail. For instance, 
in large homogeneous binary multiplicative attribute graph (MAG) models introduced by Kim and Leskovec \cite{KimLeskovec},
although (\ref{eq:ConvergenceToLittleP}) occurs, it does so only with a trivial limiting pmf $\myvec{p}$,
say $D=0$ a.s. or $D=\infty$ a.s. depending on the parameter values;
see \cite{QuMakowski} for an extended discussion.

In the homogeneous setting, the motivating issue driving the discussion is whether under Assumptions  \ref{ass:A} and  \ref{ass:B},
the convergence
\begin{equation}
P_n(d)  \probconverge_n ~ p(d),
\quad d=0,1, \ldots
\label{eq:ConvergenceFractionNodes}
\end{equation}
takes place where the pmf $\myvec{p} = ( p(d), \ d =0,1, \ldots )$
is the pmf postulated in Assumption \ref{ass:B}.
The next assumption turns out to be key.

\begin{assumption}
{\sl (Asymptotic uncorrelatedness) 
Assume that Assumption \ref{ass:A} holds, and that for each $d=0,1, \ldots $,
the identically distributed  rvs $\1{ D_{n,1} = d }$ and $\1{ D_{n,2} = d }$ are asymptotically uncorrelated
in the sense that
\begin{equation}
\lim_{n \rightarrow \infty}
{\rm Cov} \left [ \1{ D_{n,1} = d }, \1{ D_{n,2} = d } \right ] = 0.
\label{eq:C}
\end{equation}
\label{ass:C}
}
\end{assumption}

Assumption \ref{ass:C} amounts to the convergence statement
 \begin{align}
 &\lim_{n \rightarrow \infty}
 \big ( \bP{ D_{n,1} = d, D_{n,2} = d }
 \nonumber \\
 & \hspace{3mm}
 -
 \bP{ D_{n,1} = d } \bP{ D_{n,2} = d }
\big ) = 0
 \label{eq:ExplicitFormAssC}
 \end{align}
for each $d=0,1, \ldots $. It is implied by the following stronger assumption which is easier to check in
practice;  see Section \ref{sec:CommonSetting} for some examples in a commonly occurring setting.

\begin{assumption}
{\sl (Pairwise asymptotic independence) 
Assume Assumptions \ref{ass:A} and \ref{ass:B} to hold. Furthermore,
the degree rvs $D_{n,1}$ and $D_{n,2}$ are asymptotically independent
in the sense that
\begin{equation}
(D_{n,1}, D_{n,2} ) \Longrightarrow_n (D_1,D_2)
\label{eq:JointConvergenceD}
\end{equation}
where $D_1$ and $D_2$ are independent $\mathbb{N}$-valued rvs,
each distributed according to the pmf $ \myvec{p} $
postulated in Assumption \ref{ass:B}.
\label{ass:D}
}
\end{assumption}

While Assumption \ref{ass:D} reads
\[
\lim_{n \rightarrow \infty}  \bP{ D_{n,1} = d, D_{n,2} = d^\prime } 
= p(d)p(d^\prime),
\quad 
d,d^\prime=0,1, \ldots
%\label{eq:JointConvergenceEquivalent}
\]
Assumption \ref{ass:C} does not require the joint convergence (\ref{eq:JointConvergenceD}) to hold.
However, if (\ref{eq:JointConvergenceD}) were known to hold (but with {\em no} further characterization of the {\em joint} limit), 
then under Assumption \ref{ass:B} it is easy to check
that Assumption \ref{ass:C} is equivalent to the  independence of the {\em binary} rvs
$\1{ D_{1} = d }$ and $\1{ D_{2} = d }$ for each $d=0,1, \ldots $.
%Indeed, the convergence  (\ref{eq:JointConvergenceD}) certainly implies
%\[
%\lim_{n \rightarrow \infty} 
%\bP{ D_{n,1} = d, D_{n,2} = d } 
%= \bP{ D_{1} = d, D_{2} = d } 
%\]
%while Assumption \ref{ass:B} gives $\lim_{n \rightarrow \infty}  \bP{ D_{n,j} = d } = \bP{ D_{j} = d } = p(d)$ for $j=1,2$.
%Condition (\ref{eq:ExplicitFormAssC}) now yields
%\begin{eqnarray}
%\bP{ D_{1} = d , D_{2} = d }
%=
%\bP{ D_{1} = d } \bP{ D_{2} = d } ,
%\label{eq:Factor}
%\end{eqnarray}
%a statement equivalent to  the independence of the {\em binary} rvs $\1{ D_{1} = d }$ and $\1{ D_{2} = d }$.
However, the lack of independence of the rvs $D_1$ and $D_2$ does not preclude 
the possibility that the rvs $\1{ D_{1} = d }$ and $\1{ D_{2} = d }$ are independent  -- It is possible to have
$\bP{ D_{1} = d , D_{2} = d } = \bP{ D_{1} = d } \bP{ D_{2} = d }$ for {\em all} $d=0,1, \ldots$ without the rvs $D_1$ and $D_2$ being independent.

\section{A little theory -- The homogeneous case}
\label{sec:HomogeneousLittleTheory}

We return to Fact \ref{fact:BasicFact}. Fix $d=0,1, \ldots $ and $n=2,3, \ldots $. 
Under Assumption \ref{ass:A}, the expressions (\ref{eq:MEAN}) and (\ref{eq:SECOND}) become
\begin{equation}
\bE{ N_n(d) } = |V_n|  \cdot \bP{ D_{n,1} = d }
\nonumber
%\label{eq:ExpectedNumber}
\end{equation}
and
\begin{eqnarray}
\bE{ N_n(d)^2 }
&=& |V_n| (|V_n| - 1 ) \cdot \bP{ D_{n,1} = d , D_{n,2} = d }  
\nonumber \\
& &
~+ |V_n|\cdot  \bP{ D_{n,1} = d } ,
\nonumber
\end{eqnarray}
respectively.
It follows that
\begin{equation}
\bE{ P_n(d) } = \bP{ D_{n,1} = d }
\label{eq:Homogeneous1}
\end{equation}
and 
\begin{eqnarray}
\lefteqn{ {\rm Var} \left [ P_n(d) \right ] }  & &
\nonumber \\
&=&
\bE{ \left ( \frac{N_n(d)}{|V_n|} \right )^2 } 
- 
\left ( \bE{ \frac{N_n(d)}{|V_n|}  }  \right )^2
\nonumber \\
&=&
\frac{ |V_n| - 1 }{ |V_n|}  \cdot \bP{ D_{n,1} = d , D_{n,2} = d }  
\nonumber \\
& &
~+ \frac{ \bP{ D_{n,1} = d } }{|V_n|} - \left ( \bP{ D_{n,1} = d } \right )^2
\nonumber \\
&=&
\frac{  {\rm Var} \left [ \1{ D_{n,1} = d } \right ]  }{|V_n|}
\label{eq:Homogeneous2} \\
& &
+~  \frac{|V_n|-1}{|V_n|} \cdot
{\rm Cov} \left [ \1{ D_{n,1} = d }, \1{ D_{n,2} = d } \right ] 
\nonumber 
\end{eqnarray}
since $ \bP{ D_{n,1} = d }  =  \bP{ D_{n,2} = d } $ under Assumption \ref{ass:A}.

Let $n$ go to infinity in (\ref{eq:Homogeneous1}) and  (\ref{eq:Homogeneous2})  
with $\lim_{n \rightarrow \infty} |V_n| = \infty$.
It is plain that $\lim_{n \rightarrow \infty} \bE{ P_n(d) } $ exists if and only if $\lim_{n \rightarrow \infty} \bP{ D_{n,1} = d }$ exists,
and that $\lim_{n \rightarrow \infty} {\rm Var} \left [ P_n(d) \right ]  $ exists if and only if 
the limit
 \begin{equation}
C(d) 
\equiv
\lim_{n \rightarrow \infty} {\rm Cov} \left [ \1{ D_{n,1} = d }, \1{ D_{n,2} = d } \right ]
\label{eq:LimitC(d)Exists}
\end{equation}
exists.
Fact \ref{fact:BasicFact} translates into the following equivalence.

\begin{proposition}
{\sl
Assume Assumption \ref{ass:A}. 
With $d=0,1, \ldots $, we have the convergence (\ref{eq:LIMIT})
for some constant $L(d)$ in $\mathbb{R}$ if and only if
\begin{equation}
\lim_{n \rightarrow \infty} \bP{ D_{n,1} = d } = L(d)
\label{eq:LimitFirst}
\end{equation}
and
\begin{equation}
\lim_{n \rightarrow \infty} {\rm Cov} \left [ \1{ D_{n,1} = d }, \1{ D_{n,2} = d } \right ] = 0.
\label{eq:LimitCovariance}
\end{equation}
}
\label{prop:Stationary+A}
\end{proposition}

Under Assumption \ref{ass:A}, Assumption \ref{ass:B} and Assumption \ref{ass:C} 
imply that the conditions (\ref{eq:LimitFirst}) (with $L(d)=p(d)$) and (\ref{eq:LimitCovariance})
hold for {\em all} $d=0,1, \ldots $, respectively.
Applying Proposition \ref{prop:Stationary+A} we then obtain the following compact conclusion.

\begin{proposition}
{\sl
Under Assumptions \ref{ass:A}-\ref{ass:C},  the convergence (\ref{eq:ConvergenceFractionNodes}) holds
for all $d=0,1, \ldots $
where the pmf $\myvec{p} = ( p(d), \ d =0,1, \ldots )$
is the pmf postulated in Assumption \ref{ass:B}.
}
\label{prop:ConvergenceEmpirical+PMF}
\end{proposition}

To formulate a converse to Proposition \ref{prop:ConvergenceEmpirical+PMF}, 
assume Assumption \ref{ass:A} to hold. The mere existence of the limit (\ref{eq:LimitFirst}) for {\em all} $d=0,1, \ldots$
does not guarantee that the  limiting values $\{ L(d), \ d =0,1, \ldots \}$ constitute a pmf on $\mathbb{N}$. 
Without any additional assumption, it only holds that  $\sum_{d=0}^\infty L(d) \leq 1$: Indeed, for each $n=2,3, \ldots $, 
we have $\sum_{d \in V} P_n(d) \leq 1$ for every finite subset $V \subseteq \mathbb{N}$, hence
$\sum_{d \in V} \bE{ P_n(d) }  \leq 1$. By virtue of (\ref{eq:Homogeneous1})
this is equivalent to $\sum_{d \in V} \bP{ D_{n,1} = d }  \leq 1$. Letting $n$ go to infinity we get
$\sum_{d \in V} p(d) \leq 1$, and the desired conclusion follows.

\begin{proposition}
{\sl
Under Assumption \ref{ass:A}, assume that for every $d=0,1, \ldots $, there exists a scalar $L(d)$ such that 
\begin{equation} 
 P_n(d)  \probconverge_n ~ L(d).
 \label{eq:ConvergenceEmpirical+Converse}
\end{equation}
If the limiting values $\{ L(d), \ d =0,1, \ldots \}$ constitute a pmf $\myvec{p} = ( p(d), \ d=0,1, \ldots )$ on $\mathbb{N}$, then both 
Assumption \ref{ass:B} (with pmf $\myvec{p}$) and Assumption \ref{ass:C} must hold.
}
\label{prop:ConvergenceEmpirical+Converse}
\end{proposition}

\myproof
By Proposition \ref{prop:Stationary+A}, the validity of  (\ref{eq:ConvergenceEmpirical+Converse}) for all $d=0,1, \ldots $
implies  (\ref{eq:LimitCovariance}) for all $d=0,1, \ldots $, hence
Assumption \ref{ass:C} holds.
By Proposition \ref{prop:Stationary+A}, the validity of (\ref{eq:ConvergenceEmpirical+Converse}) for all $d=0,1, \ldots $
also implies that (\ref{eq:LimitFirst}) holds for all $d=0,1, \ldots$.
If additionally the  limiting values $\{ L(d), \ d =0,1, \ldots \}$ constitute a pmf $\myvec{p}$ on $\mathbb{N}$, then there exists
an $\mathbb{N}$-valued rv $D$ distributed according to the pmf $\myvec{p}$ such that $D_{n,1} \Longrightarrow_n D$, and Assumption \ref{ass:B} holds.
\myendpf

As we survey the discussion so far, it is plain that under Assumptions \ref{ass:A}-\ref{ass:B},
the convergence (\ref{eq:LIMIT}) 
necessarily takes the form (\ref{eq:ConvergenceFractionNodes}). Furthermore, whenever we have
\begin{equation}
\lim_{n \rightarrow \infty}
{\rm Cov} \left [ \1{ D_{n,1} = d }, \1{ D_{n,2} = d } \right ]  > 0,
\label{eq:LimInfCondition1}
\end{equation}
then (\ref{eq:ConvergenceFractionNodes}) cannot hold.

\section{A little theory -- The general setting}
\label{sec:GeneralSettingLittleTheory}

Proposition \ref{prop:Stationary+A}  is a special case of a more general fact that does not require any homogeneity assumption.
We devote this section to a presentation of this more general viewpoint:
Fix $n=2,3, \ldots $.
In the context of the random graph $\mathbb{G}_n$, let $\Sigma_n$
denote the set $\{ (k,\ell) \in V_n \times V_n :  \ k \neq \ell \}$ that comprises all ordered pairs drawn from $V_n$
without repetition.
Let also the rv $(\nu_n,\mu_n): \Omega \rightarrow \Sigma_n$ be {\em uniformly} distributed over  $\Sigma_n$, i.e.,
\[
\bP{ \nu_n= k, \mu_n = \ell }
= \frac{1}{|V_n|(|V_n|-1)},
\quad
\begin{array}{c}
k \neq \ell \\
k,\ell \in V_n. \\
\end{array}
\]
Thus, the rv $(\nu_n,\mu_n)$ models the randomly uniform selection of two nodes in $V_n$ (without repetition); 
the rvs $\nu_n$ and $\mu_n$ are both {\em uniformly} distributed over $V_n$.
The selection rv $(\nu_n,\mu_n)$ is assumed to be {\em independent} of the random graph $\mathbb{G}_n$.

Fix $d=0,1, \ldots $. Under the enforced independence assumptions, we note from (\ref{eq:MEAN}) that
\begin{eqnarray}
\bE{ \frac{N_n(d)}{|V_n|} } 
&=&  \frac{1}{|V_n|} \cdot \sum_{k \in V_n }  \bP{ D_{n,k} = d }
\nonumber \\
&=& \sum_{k \in V_n } \bP{ \nu_n = k, D_{n,k} = d },
\nonumber
\end{eqnarray}
and it follows that
\begin{equation}
\bE{ P_n(d) }  = \bP{ D_{n,\nu_n} = d } .
\label{eq:AAA}
\end{equation}
Using  (\ref{eq:SECOND}) we also conclude from (\ref{eq:AAA}) that
\begin{eqnarray}
\lefteqn{ \bE{ N_n(d)^2 } } & &
\nonumber \\
&=& |V_n |  \cdot \bP{ D_{n,\nu_n} = d }
\nonumber \\
& &
+~ 
|V_n| \left ( |V_n| - 1 \right ) \cdot  \bP{ D_{n,\nu_n} = d , D_{n,\mu_n} = d  },
\nonumber
%\label{eq:CAA}
\end{eqnarray}
whence
\begin{eqnarray}
\lefteqn{
{\rm Var} \left [  P_n(d) \right ]
} & &
\nonumber \\
&=& \bE{ \left ( \frac{N_n(d)}{|V_n|} \right ) ^2 }
- 
\left ( \bE{  \frac{N_n(d)}{|V_n|} }  \right )^2
\nonumber \\
&=& \frac{ \bP{ D_{n,\nu_n} = d } }{|V_n|} + \frac{|V_n|-1}{|V_n|}  \cdot \bP{ D_{n,\nu_n} = d , D_{n,\mu_n} = d  }
\nonumber \\
& &  -~ \bP{ D_{n,\nu_n} = d } \bP{ D_{n,\mu_n} = d  } 
\nonumber \\
&=& \frac{ {\rm Var } \left [ \1{ D_{n,\nu_n} = d  } \right ]  }{ |V_n|} 
\label{eq:DAA} \\
& &
+~ \frac{ |V_n|-1}{|V_n|} \cdot {\rm Cov} \left [ \1{ D_{n,\nu_n} = d } , \1{ D_{n,\mu_n} = d  } \right ].
\nonumber
\end{eqnarray}
To obtain the variance term in (\ref{eq:DAA}) we used the obvious equality
$\bP{ D_{n,\mu_n} = d } = \bP{ D_{n,\nu_n} = d }$.
Let $n$ go to infinity in (\ref{eq:AAA}) and (\ref{eq:DAA})
with $\lim_{n \rightarrow \infty} |V_n|=\infty$. 
Appealing again to Fact \ref{fact:BasicFact} we obtain the following analog of Proposition \ref{prop:Stationary+A}.

\begin{proposition}
{\sl
Under the foregoing assumptions, with $d=0,1, \ldots $, the convergence (\ref{eq:LIMIT}) holds
for some scalar $L(d)$ in $\mathbb{R}$ if and only if 
\begin{equation}
\lim_{n \rightarrow \infty }  \bP{ D_{n,\nu_n} = d } = L(d)
\label{eq:LBAA}
\end{equation}
and
\begin{equation}
\lim_{n \rightarrow \infty }  {\rm Cov} \left [ \1{ D_{n,\nu_n} = d } , \1{ D_{n,\mu_n} = d  } \right ] = 0 .
\label{eq:LBBA}
\end{equation}
}
\label{prop:GeneralSetting}
\end{proposition}

Under Assumption \ref{ass:A}, for each $n=1,2, \ldots $ the distributional equalities
$D_{n,\nu_n} =_{st} D_{n,1}$, $D_{n,\mu_n} =_{st} D_{n,1}$,
and $( D_{n,\nu_n}, D_{n,\mu_n} )=_{st} ( D_{n,1}, D_{n,2})$ hold, in which case the
conditions (\ref{eq:LBAA}) and (\ref{eq:LBBA}) reduce to
conditions (\ref{eq:LimitFirst}) and (\ref{eq:LimitCovariance}) of Proposition \ref{prop:Stationary+A}, respectively --
Proposition \ref{prop:Stationary+A} is plainly subsumed by Proposition \ref{prop:GeneralSetting}.
While the latter holds under {\em no} assumption on the sequence $\{ \mathbb{G}_n , \ n=2,3, \ldots \}$, unfortunately in that generality it does
not retain the operational ability of Proposition \ref{prop:Stationary+A} of {\em equating} the two different degree distributions
available in the homogeneous case.

Proposition \ref{prop:GeneralSetting} also applies when the graphs $\{ \mathbb{G}_n, \ n=2,3, \ldots \}$ are deterministic.
The {\em non}-homogeneous Barab\'asi-Albert model
(and other growth models) are easily  fitted into this more general framework. 
In particular, Proposition \ref{prop:GeneralSetting}  offers the possibility
of establishing the convergence (\ref{eq:LIMIT}) through (\ref{eq:LBAA}) and (\ref{eq:LBBA}). 
These two properties follow if the sequence $\{ \mathbb{G}_n, \ n=2,3, \ldots \}$ is {\em locally weakly convergent}
(or weakly convergent in the sense of Benjamini and Schramm \cite{BenjaminiSchramm}); see
the reference \cite{AldousSteele} for an introduction to these ideas.
The Barab\'asi-Albert model (and some of its variants) were shown to be locally weakly convergent by Berger et al. 
\cite{BergerBorgsChayesSaberi}.
However, in the Barab\'asi-Albert model, Bollob\'{a}s et al. have shown the convergence (\ref{eq:LIMIT})
by direct Hoeffding-Azuma bounding arguments \cite{Bollobas_BAmodel},
thereby implying (\ref{eq:LBBA}) (as well as (\ref{eq:LBAA}) trivially by bounded convergence).

\section{Convergence in total variation distance}
\label{sec:Totalvariation}

The weak convergence of $\mathbb{N}$-valued rvs is equivalent to convergence 
in the total variation distance of their corresponding pmfs (on $\mathbb{N}$); this is a well-known consequence
of Scheff\'e's Theorem \cite[App. II, p. 224]{BillingsleyBook} when applied to discrete rvs. Here we show an analogous 
equivalence when the convergence  in {\em probability} (\ref{eq:ConvergenceFractionNodes}) holds for all $d=0,1, \ldots$:

If $\myvec{\mu} = ( \mu(x), \ x=0,1, \ldots )$ and $\myvec{\nu} = ( \nu(x), \ x=0,1, \ldots )$ are two
pmfs on $\mathbb{N}$, the {\em total variation distance} between them is given by
\[
d_{\rm TV} ( \myvec{\mu}, \myvec{\nu} )
=
\frac{1}{2} \sum_{x=0}^\infty \left | \mu(x) - \nu(x) \right |.
\]
This quantity can alternatively be expressed as
\[
d_{\rm TV} ( \myvec{\mu}, \myvec{\nu} )
= \sum_{x=0}^\infty  \left ( \mu(x) - \nu(x) \right )^+ = \sum_{x=0}^\infty  \left ( \nu(x) - \mu(x) \right )^+ .
%\label{eq:TotalRepresenationRepresentation}
\]

\begin{proposition}
{\sl
Under Assumptions \ref{ass:A}-\ref{ass:C},  we have
\begin{equation}
d_{\rm TV}
\left ( \myvec{P}_n , \myvec{p} \right ) 
\probconverge_n  ~ 0 
\label{eq:TotalVariationConvergenceInProbability}
\end{equation}
where the pmf $\myvec{p} = ( p(d), \ d =0,1, \ldots )$
is the pmf postulated in Assumption \ref{ass:B}.
}
\label{prop:ConvergenceEmpirical+TotalVariation}
\end{proposition}

\myproof
Pick $V \subseteq \mathbb{N}_0$ arbitrary with $|V| < \infty$. 
Using the alternate representation above, we get
\begin{eqnarray}
\bE{ d_{\rm TV} \left ( \myvec{P}_n ,  \myvec{p} \right )  }
\leq
\sum_{x \in V}  \bE{  \left ( p(x) - P_n(x)  \right )^+ }
+ \sum_{x \notin V} p(x)
\nonumber 
\end{eqnarray}
for each $n=1,2, \ldots $. Let $n$ go to infinity in this last inequality.
By Proposition \ref{prop:ConvergenceEmpirical+PMF} we have $P_n (x)  \probconverge_n ~ p(x)$ for each $x=0,1, \ldots$, whence
$\lim_{n \rightarrow \infty} \bE{ \left ( p(x) - P_n(x)  \right )^+  } = 0$ by bounded convergence.
It follows that
\[
\limsup_{n \rightarrow \infty}  \bE{ d_{\rm TV} \left ( \myvec{P}_n ,  \myvec{p} \right )  } \leq \sum_{x \notin V} p(x). 
\]
The set $V$ being an arbitrary finite subset of $\mathbb{N}$ and $\myvec{p}$ being a pmf on $\mathbb{N}$ (hence tight), we readily obtain
$\lim_{n \rightarrow \infty} \bE{ d_{\rm TV} \left ( \myvec{P}_n ,  \myvec{p} \right )  }  = 0$,
and the desired conclusion 
(\ref{eq:TotalVariationConvergenceInProbability}) follows by Markov's inequality.
\myendpf

\section{A commonly encountered setting}
\label{sec:CommonSetting}

In many situations of interest the sequence of random graphs  $\{ \mathbb{G}_n, \ n=1,2, \ldots \}$
arises in the following natural manner:
Given is an {\em underlying} parametric family of random graphs, say
\begin{equation}
\{ \mathbb{G}(n;\alpha), \ n=2,3, \ldots \},
\quad \alpha \in A \subseteq \mathbb{R}^r
\label{eq:UnderlyingFamily}
\end{equation}
where $A$ is some parameter set and $r$ is a positive integer.
With $\alpha$ in $A$,  for each $n=2,3, \ldots $, the random graph
$\mathbb{G}(n;\alpha)$ is a random graph on $V_n$ whose statistics  depend on the parameter $\alpha$.
For each $k$ in $V_n$, let $D_{n,k}(\alpha)$ denote the degree of node $k$ in $\mathbb{G}(n;\alpha)$;
it is often the case that the rvs $\{ D_{n,k} (\alpha) , \ k \in V_n \}$ constitute an {\em exchangeable} family,
as we assume thereafter in this section.
Thus, there is no ambiguity when speaking of  {\em the} (nodal) degree distribution in $\mathbb{G}(n;\alpha)$
because {\em all} nodes have the same degree distribution, namely that of the rv $D_{n,1} (\alpha)$.

We construct the collection $\{ \mathbb{G}_n, \ n=2,3, \ldots \}$ by setting
\begin{equation}
\mathbb{G}_n \equiv \mathbb{G}(n;\alpha_n),
\quad n=2,3, \ldots
\label{eq:SettingWithUnderlyingFamily}
\end{equation}
for some scaling $\alpha: \mathbb{N}_0 \rightarrow A$, in which case $D_{n,k} = D_{n,k} (\alpha_n)$ for each $k$ in $V_n$ --
Scalings are sequences which we view as mappings defined on $\mathbb{N}_0$; the mapping itself is denoted by
the same symbol used for the generic element of the sequence.

The scaling $\alpha: \mathbb{N}_0 \rightarrow A$ appearing in (\ref{eq:SettingWithUnderlyingFamily})
is the (usually unique) scaling which ensures the convergence
\begin{equation}
D_{n,1}(\alpha_n) \Longrightarrow_n D
\label{eq:ConvergenceToD}
\end{equation}
for some non-degenerate $\mathbb{N}$-valued rv $D$; 
this scaling is often the critical scaling associated with the emergence of a maximal component.
Under these circumstances, Assumptions \ref{ass:A} and \ref{ass:B} are automatically satisfied,
and only Assumption \ref{ass:C} needs to be verified.

The setting outlined above applies to a number of examples routinely
discussed in the literature: Here for each $n=1,2, \ldots $, we take $V_n = \{1, \ldots , n \}$. With $c > 0$,

\begin{enumerate}
\item
Erd\H{o}s-R\'enyi graphs $\mathbb{G}(n;p)$ ($0 \leq p \leq 1$)
with scaling $p: \mathbb{N}_0 \rightarrow [0,1]$ such that $p_n \sim \frac{c}{n}$
\cite{Bollobas_Book,DraiefMassoulieLMMS,ER1960};
\item
Geometric random graphs $\mathbb{G}(n;\rho)$ on a unit square 
($\rho > 0$) with scaling $\rho: \mathbb{N}_0 \rightarrow \mathbb{R}_+ $ such that $ \pi \rho_n^2 \sim \frac{c}{n}$
\cite{PenroseBook}; and
\item
Random key graphs $\mathbb{K}(n;K,P)$ ($K < P$ in $\mathbb{N}_0$)
with scalings $K,P: \mathbb{N}_0 \rightarrow \mathbb{N}_0$ such that $\frac{K^2_n}{P_n}  \sim \frac{c}{n}$
\cite{YaganMakowskiConnectivity}.
\end{enumerate}

Assumption \ref{ass:A} is readily satisfied in these homogeneous situations.
In each case, Poisson convergence can be invoked to validate Assumption \ref{ass:B} with  the rv $D$ in  (\ref{eq:ConvergenceToD})
being a Poisson rv with parameter $c$. 
In all cases, the stronger Assumption \ref{ass:D} is established, thereby implying Assumption \ref{ass:C}. 
While it is  elementary to do so for Erd\H{o}s-R\'enyi graphs,
the calculations become increasingly tedious as we move from geometric random graphs to random key graphs;
see \cite{SPal_Thesis} for details.
Finally, despite an abundance of situations where Assumptions \ref{ass:A}-\ref{ass:C} are satisfied (beyond the ones discussed above),
it is nevertheless possible to find homogeneous random networks in the sense of Assumption \ref{ass:A}
where (\ref{eq:ConvergenceToLittleP}) occurs but where the convergence (\ref{eq:ConvergenceFractionNodes}) fails. 
This is taken on in the remainder of the paper starting with the next section.

\section{A counterexample} 
\label{sec:CounterExample}

\subsection{Random threshold graphs}

The setting is that of \cite{CCDM,  FIKMMU, MakowskiYagan-JSAC, SC}:
Let $\{ \xi, \xi_k, \ k=1,2, \ldots \}$
denote a collection of i.i.d. $\mathbb{R}_+$-valued rvs defined on 
the probability triple $(\Omega, {\cal F}, \mathbb{P})$,
each distributed according to a given
(probability) distribution function 
$F: \mathbb{R} \rightarrow [0,1]$ with
$F(x) = 0$ for $x\leq 0$.
With $\xi$ acting as  a generic representative for this sequence of i.i.d. rvs,
we have
\[
\bP{ \xi \leq x } = F(x),
\quad x \in \mathbb{R}.
\]

Once $F$ is specified, random thresholds graphs are characterized by two parameters, 
namely a positive integer $n$ and a threshold value $\theta > 0$:
The network comprises $n$ nodes, labelled $k=1, \ldots , n$,
and to each node $k$ we assign a {\em fitness} variable 
(or weight) $\xi_k$ which measures its importance or rank.
For distinct $k,\ell =1, \ldots , n$,
the nodes $k$ and $\ell$ are declared to be adjacent if
\begin{equation}
\xi_k + \xi_\ell > \theta ,
\label{eq:AdjacencyDefnRTGs}
\end{equation} 
and a bidirectional edge exists between nodes $k$ and $\ell$. 
The adjacency notion (\ref{eq:AdjacencyDefnRTGs}) defines the {\em random threshold} graph 
$\mathbb{T}(n;\theta)$ on the set of vertices $V_n = \{ 1, \ldots , n \}$.
The degree $D_{n,k}(\theta)$ of node $k$ in $\mathbb{T}(n;\theta)$ is clearly given by
\[
D_{n,k}(\theta) = \sum_{\ell=1, \ \ell \neq k}^n \1{ \xi_k + \xi_\ell > \theta },
\quad k=1, \ldots , n.
\]
Under the enforced assumptions, the rvs $D_{n,1}(\theta), \ldots , D_{n,n}(\theta)$
are exchangeable, thus equidistributed.

\subsection{Applying Proposition \ref{prop:ConvergenceEmpirical+PMF} under exponential fitness}

From now on we focus on the special case 
when $\xi$ is exponentially distributed
with parameter $\lambda > 0$, written $\xi \sim {\rm Exp}(\lambda)$,  that is
\begin{equation}
\bP{ \xi \leq x } 
= 1 - e^{-\lambda x^+},
\quad x \in \mathbb{R}
\label{eq:CDF=Exponential}
\end{equation}
where we have used the standard notation $x^+ = \max(x,0)$.
While other distributions could be considered to develop counterexamples to Proposition \ref{prop:ConvergenceEmpirical+PMF},
the exponential distribution was selected for two main reasons:
This situation was considered in the references \cite{CCDM, FIKMMU, SC}
in making the case that scale-free networks can be generated through the fitness-based mechanism used
in random threshold graphs; more on that later.
Moreover, calculations are greatly simplified in the exponential case.

With random threshold graphs as the underlying family (\ref{eq:UnderlyingFamily}), 
the definition (\ref{eq:SettingWithUnderlyingFamily}) here takes the form
\begin{equation}
\mathbb{G}_n = \mathbb{T}(n; \theta^\star_n),
\quad n=2,3, \ldots 
\label{eq:G=RTG}
\end{equation}
with scaling $\theta^\star : \mathbb{N}_0 \rightarrow [0,\infty)$ given by
\begin{equation}
\theta^\star _n = \lambda^{-1} \log n,
\quad n=2, 3, \ldots .
\label{eq:FujiharaScaling}
\end{equation}

We are in the setting of Section \ref{sec:CommonSetting}.
Having in mind to apply Proposition \ref{prop:ConvergenceEmpirical+PMF} to the random graphs
$\{ \mathbb{G}_n, \ n=1,2, \ldots \}$,  we recover the notation of Section \ref{sec:Framework} by setting
\[
D_{n,k} = D_{n,k}(\theta^\star_n),
\quad 
\begin{array}{c}
k=1, \ldots , n\\
n=2,3,\ldots \\
\end{array}
\]
Assumption \ref{ass:A} is obviously satisfied in light of the aforementioned exchangeability.
It was shown by Fujihara et al. \cite[Example 1, p. 366]{FIKMMU} that
$D_{n,1} \Longrightarrow_n D$
where the $\mathbb{N}$-valued rv $D$ is a conditionally Poisson rv with pmf 
$\myvec{p}_{\rm Fuj} = ( p_{\rm Fuj} (d) , \ d=0,1, \ldots )$ given by
\begin{eqnarray}
p_{\rm Fuj} (d) 
= \bP{ D=d } 
= \bE{ \frac{ (e^{\lambda \xi})^d }{d!} e^{-e^{\lambda \xi}} },
\quad d=0,1, \ldots
\label{eq:Fujihara}
\end{eqnarray}
Therefore, Assumption \ref{ass:B} holds with
\begin{equation}
\lim_{n \rightarrow \infty} \bP{ D_{n,1} =d } = p_{\rm Fuj} (d) ,
\quad d=0,1, \ldots
\label{eq:ConvergenceToFuj}
\end{equation}

\subsection{Assumption \ref{ass:C} fails}

The remainder of the paper is devoted to showing the following convergence result.

\begin{proposition}
{\sl
Assume $\xi \sim {\rm Exp}(\lambda)$ for some $\lambda > 0$. For each $d=0,1, \ldots $, the limit
\begin{align}
C(d)
\equiv 
\lim_{n \rightarrow \infty}
{\rm Cov} \left [ \1{ D_{n,1} (\theta^\star_n) = d } ,  \1{ D_{n,2} (\theta^\star_n) = d } \right ]
\label{eq:AssumptionC+fails}
\end{align}
exists and $C(d) > 0$.
}
\label{prop:AssumptionC+fails}
\end{proposition}

Proposition \ref{prop:AssumptionC+fails} is established from Section \ref{sec:ProofPart+1} to Section \ref{sec:ProofPart+2} where
expressions are given for the limits (\ref{eq:AssumptionC+fails}): 
For instance,  we show at (\ref{eq:C(0)}) that
\[
C(0)
=
\bE{ e^{ -\max( e^{\lambda \xi_1}, e^{\lambda \xi_2} ) }  }  - \bE{ e^{ - ( e^{ \lambda \xi_1  }  + e^{ \lambda  \xi_2  } ) }  } > 0.
\]
The expression (\ref{eq:C(d)}) of the limit $C(d) $ for $d \neq 0$ is rather cumbersome and is omitted at this point.
However, the fact that $C(d) > 0$ on the entire range suffices to establish the desired counterexample
by virtue of the observation following Proposition \ref{prop:ConvergenceEmpirical+Converse}.

\begin{proposition}
{\sl Assume $\xi \sim {\rm Exp}(\lambda)$ for some $\lambda > 0$. For each $d=0,1, \ldots $, the sequence of rvs
\begin{equation}
\left \{
\frac{1}{n}
\sum_{k=1}^n \1{ D_{n,k} (\theta^\star_n) = d },
\ n= 2, 3, \ldots 
\right \}
\label{eq:NonConvergenceInProbability}
\end{equation}
does not converge in probability to any constant.
}
\label{prop:NonConvergenceInProbability}
\end{proposition}

In fact, for each $d=0,1, \ldots $, there exists
a non-degenerate $[0,1]$-valued rv $\Pi(d)$ with $\bE{ \Pi (d) } = p_{\rm Fuj}(d)$ and ${\rm Var} \left [ \Pi(d) \right ] > 0$
such that
\[
\frac{1}{n}
\sum_{k=1}^n \1{ D_{n,k} (\theta^\star_n) = d }  \Longrightarrow_n ~ \Pi(d).
\]
Details are available in \cite{SPal_Thesis}.

The failure of the convergence (\ref{eq:ConvergenceFractionNodes}) in the
context of random threshold graphs with exponentially distributed fitness is noteworthy for the following reason:
Caldarelli et al. \cite{CCDM,SC} have proposed this class of random graph models
as an alternative scale-free model to the preferential attachment model of Barab\'{a}si and Albert \cite{BarabasiAlbert}.
The basis for their proposal was the provable power-law behavior
\begin{equation}
p_{\rm Fuj}(d) \sim d^{-2}
\quad (d \rightarrow \infty).
\label{eq:TailFujihara}
\end{equation}
See Fujihara et al. \cite[Example 1, p. 366]{FIKMMU} for details.
However,  a meaningful comparison between the two models would have required at  minimum
the validity of the convergence
\[
\frac{1}{n}
\sum_{k=1}^n \1{ D_{n,k} (\theta^\star_n) = d }  \probconverge_n ~p_{\rm Fuj} (d) ,
\quad d=0,1, \ldots 
\]
By Proposition \ref{prop:NonConvergenceInProbability}
this last convergence fails to happen, and the two models cannot be meaningfully compared
since  for the Barab\'{a}si-Albert model it only holds that
\[
\frac{1}{n}
\sum_{k=1}^n \1{ D_{n,k} (\theta^\star_n) = d } \probconverge_n ~ p_{\rm BA} (d), 
\quad d=0,1, \ldots 
\]
with $p_{\rm BA}(d) \sim d^{-3}$ $(d \rightarrow \infty)$  \cite{Bollobas_BAmodel}.
Although the Barab\'asi-Albert model has attracted much attention as a network model, 
its tree-like structure does not make it a particularly good  fit for the empirical data coming from large real-life networks.
Similar comments apply to the class of random threshold graph models due to a propensity to produce star-like structures.

\section{Simulation Results}
\label{sec:SimulationResults}

Through a limited set of simulation experiments, we now demonstrate the failure of the convergence
(\ref{eq:ConvergenceFractionNodes}) established for random threshold graphs in Proposition \ref{prop:NonConvergenceInProbability}. 
Throughout, the fitness variable $\xi$ is taken to be exponentially distributed with parameter $\lambda=1$, 
and the threshold is scaled in accordance to (\ref{eq:FujiharaScaling}), namely $\theta _n ^\star = \log n$ for each $n=2,3,\ldots$.

The number $n$ of nodes being given, 
we generate $R$ mutually independent realizations of the random threshold graph $\mathbb{T}(n;\theta _n ^\star)$; 
they are denoted $\mathbb{T}^{(1)}(n;\theta_n^\star), \bT^{(2)}(n;\theta_n^\star), \ldots, \mathbb{T}^{(n)}(n;\theta _n ^\star)$, respectively.
For each $k=1,2,\ldots,n$ and $r=1,2,\ldots,R$, let $D_{n,k} ^{(r)}(\theta _n^\star)$ denote the degree of node $k$ in the random graph 
$\mathbb{T}^{(r)}(n;\theta_n^\star)$.

\begin{figure}[h]
\centering
\subfloat[$n=10000$, $R=100$\label{fig:Stripplot_10000nodes]}]{\includegraphics[width=\linewidth]{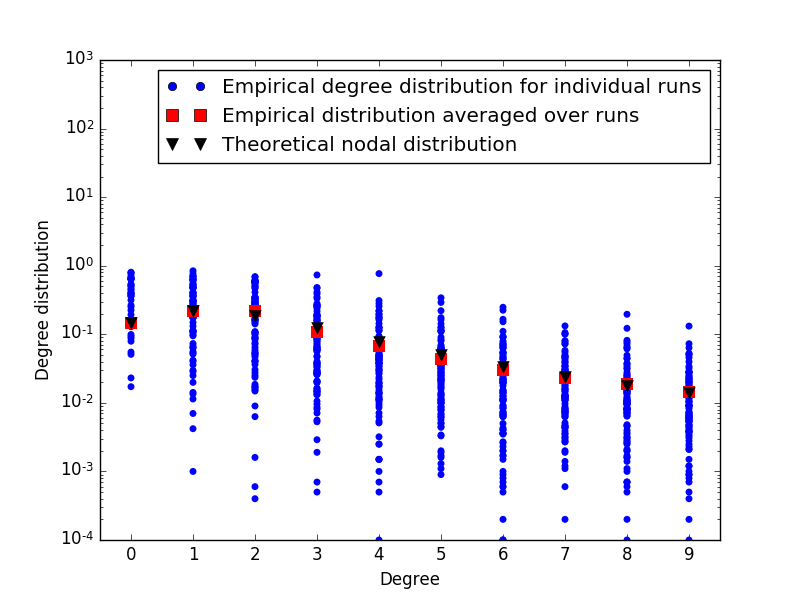}}
 \hfill
\subfloat[$n=30000$, $R=100$\label{fig:Stripplot_30000nodes}]{\includegraphics[width=\linewidth]{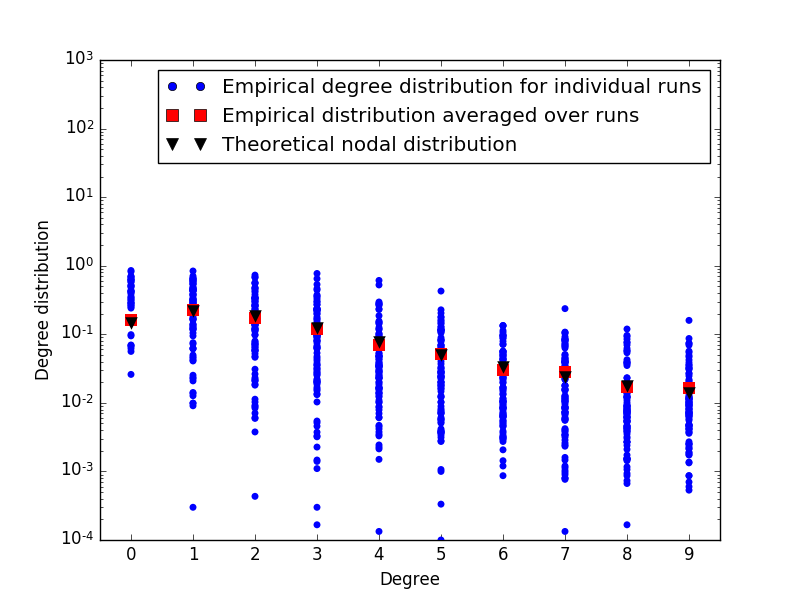}}
\caption{}%The nodal degree distribution $\myvec{p}_{\rm Fuj}$  was plotted against the  empirical degree distribution 
               % $\frac{N_n  ^{(r)}( . ;\theta _n ^{\star})}{n}$ for various runs $r=1,2,\ldots,R$. }
\label{fig:Stripplot_EmpiricalVsNodal}
\end{figure}
We explore the behavior of the empirical degree distribution along the scaling (\ref{eq:FujiharaScaling})
(with $\lambda=1$) as generated through a single network realization. We do so by plotting the histograms 
\begin{equation}
\frac{N_n ^{(r)}(d;\theta_n ^\star)}{n} = \frac{1}{n} \sum_{k=1}^n \1{ D^{(r)}_{n,k} (\theta^\star_n) = d },
\quad 
\begin{array}{c}
d=0,1,\ldots \\
r=1, \ldots , R \\
\end{array}
\label{eq:SingleRealization}
\end{equation}
for various values of $d$ and $r$, and large $n$, and comparing against the corresponding value for
the limiting nodal distribution $p_{\rm Fuj} (d)$ given in (\ref{eq:Fujihara}).
Using this expression we numerically evaluate $p_{\rm Fuj} (d)$ as
\begin{equation}
p_{\rm Fuj} (d)  
= \frac{1}{d!}  \int_0^\infty  \frac{e^{d x} }{d!} e^{-e^{x} } e^{-x} dx
\nonumber
\end{equation}
for each $ d=0,1, \ldots$.
In Figure \ref{fig:Stripplot_EmpiricalVsNodal} we plot the histogram $\frac{N_n ^{(r)}( . ;\theta_n ^\star)}{n}$ 
for different runs $r=1,2,\ldots,R$ and varying graph sizes $n=10000,30000$. Observe the high variability 
with respect to the nodal degree distribution $\myvec{p}_{\rm Fuj}$ which does not change as the graph size is increased.

We smooth out the variability observed in Figure \ref{fig:Stripplot_EmpiricalVsNodal}
by averaging the empirical degree distributions (\ref{eq:SingleRealization}) over the $R$ i.i.d. realizations
$\mathbb{T}^{(1)}(n; \theta ^\star _n),  \mathbb{T}^{(2)}(n; \theta ^\star _n) ,  \ldots ,  \mathbb{T}^{(R)}(n; \theta ^\star _n)$.
This results in the statistic
\begin{equation}
\frac{1}{R} \sum _{r=1} ^R \frac{N_n ^{(r)}(d;\theta_n ^\star)}{n}, 
\ d=0,1,\ldots .
\label{eq:Run_averaged_empirical_distribution}
\end{equation}
Fix $d=0,1, \ldots $. The Strong Law of Large Numbers yields
\begin{eqnarray}
\lim_{R \rightarrow \infty} \frac{1}{R} \sum _{r=1} ^R \frac{N_n ^{(r)}(d;\theta_n ^\star)}{n}
= \bE{ \frac{N_n (d;\theta_n ^\star)}{n} }
\quad \mbox{a.s.}
\end{eqnarray}
with
\[
 \bE{ \frac{N_n (d;\theta_n ^\star)}{n} } = \bP{  D_{n,1} (\theta^\star_n) = d }
\]
by exchangeability. On the other hand, by virtue of (\ref{eq:ConvergenceToFuj}) we have
$\lim_{n \rightarrow \infty} \bP{  D_{n,1} (\theta^\star_n) = d } = p_{\rm Fuj}(d)$.
Combining these observations yields the approximation
\begin{equation}
\frac{1}{R} \sum _{r=1} ^R \frac{N_n ^{(r)}(d;\theta_n ^\star)}{n} =_{\rm Approx} p_{\rm Fuj}(d)
\label{eq:Approx4}
\end{equation}
for large $n$ and $R$. 
The goodness of the approximation (\ref{eq:Approx4}) is noted in Figure \ref{fig:Stripplot_EmpiricalVsNodal}, 
where the empirical distribution averaged over $R=100$ runs is observed to be very close to the nodal degree distribution.
However, the accuracy of the approximation (\ref{eq:Approx4}) does in no way imply the validity of  (\ref{eq:ConvergenceFractionNodes}). 
In fact the mistaken belief that (\ref{eq:ConvergenceFractionNodes}) holds,
implicitly assumed in the papers \cite{CCDM,SC}, might have stemmed from using the smoothed estimate (\ref{eq:Approx4}).

\section{Preparing the proof of Proposition \ref{prop:AssumptionC+fails}}
\label{sec:ProofPart+1}

For every $n=2,3, \ldots $ and $\theta > 0$, the decomposition
\begin{equation}
D_{n,j}(\theta)  = \1{ \xi_1 + \xi_2 > \theta } + D^\star_{n,j}(\theta),
\quad j=1,2
\label{eq:DegreeDecomposition}
\end{equation}
holds where we have set
\[
D^\star_{n,j}(\theta) 
= \sum_{k=3}^n \1{ \xi_j + \xi_k > \theta }.
\]

Fix $d=0,1, \ldots $. It is a simple matter to check that
\begin{eqnarray}
\lefteqn{ \left | \bP{ D_{n,j} (\theta) = d }
-
\bP{ D^\star_{n,j} (\theta) = d  }
\right | } & &
\nonumber \\
&\leq& 2 \bP{ \xi_1 + \xi_2 > \theta } ,
\quad j=1,2
\label{eq:BoundOnDifference1}
\end{eqnarray}
and
\begin{eqnarray}
& &
\left | \bP{ D_{n,j} (\theta) = d , \ j=1,2 } - \bP{ D^\star_{n,j} (\theta) = d , \ j=1,2 }  \right | 
\nonumber \\
& & \leq 2 \bP{ \xi_1 + \xi_2 > \theta } .
\label{eq:BoundOnDifference2}
\end{eqnarray}
Next, for each $n=2,3, \ldots$ we substitute $\theta$ by $\theta^\star_n$ in 
the bound (\ref{eq:BoundOnDifference2}) %according to (\ref{eq:FujiharaScaling}),
and let $n$ go to infinity in the resulting inequality.
Since $\lim_{n \rightarrow \infty} \theta^\star_n = \infty$, we conclude that
$ \lim_{n \rightarrow \infty} 
\big | \bP{ D_{n,1} (\theta^\star_n) = d , D_{n,2} (\theta^\star_n) = d  }
-
\bP{ D^\star_{n,1} (\theta^\star_n) = d , D^\star_{n,2} (\theta^\star_n) = d  }
\big | =0$,
whence
\begin{eqnarray}
& &
\lim_{n \rightarrow \infty} 
\bP{ D_{n,1} (\theta^\star_n) = d , D_{n,2} (\theta^\star_n) = d  }
\nonumber \\
&=&
\lim_{n \rightarrow \infty} 
\bP{ D^\star_{n,1} (\theta^\star_n) = d , D^\star_{n,2} (\theta^\star_n) = d  }
\label{eq:TwoDimConvergence}
\end{eqnarray}
provided either limit exists. The same argument  applied to the bounds
(\ref{eq:BoundOnDifference1}) readily yields
\begin{align}
\lim_{n \rightarrow \infty} 
\bP{  D_{n,j} (\theta^\star_n) = d } 
&=
\lim_{n \rightarrow \infty} 
\bP{  D^\star _{n,j} (\theta^\star_n) = d } 
\nonumber \\
&= p_{\rm Fuj}(d) ,
\quad j=1,2
\label{eq:OneDimConvergence}
\end{align}
in light of (\ref{eq:ConvergenceToFuj}).
It then follows from
(\ref{eq:TwoDimConvergence}) and  (\ref{eq:OneDimConvergence}) that
\begin{eqnarray}
\lefteqn{ C(d) } & &
\label{eq:LimitGivesC(d)} \\
&=& \lim_{n \rightarrow \infty}
{\rm Cov} \left [ \1{ D^\star_{n,1} (\theta^\star_n) = d } ,  \1{ D^\star_{n,2} (\theta^\star_n) = d } \right ] 
\nonumber
\end{eqnarray}
provided either limit at
(\ref{eq:TwoDimConvergence}) exists.

As we now turn to evaluating (\ref{eq:LimitGivesC(d)}), it will be notationally
convenient to introduce a second collection of $\mathbb{R}_+$-valued rvs
$\{ \eta_\ell, \ \ell=1,2, \ldots \}$. We assume that the rvs $\{ \eta_\ell, \ \ell=1,2, \ldots \}$ are also i.i.d. rvs, 
each of which is exponentially distributed with parameter $\lambda > 0$.
The two collections $\{ \xi, \xi_k, \ k=1,2, \ldots \}$
and $\{ \eta_\ell, \ \ell =1,2, \ldots \}$ are assumed to be mutually independent.
For each integer $p=2,3, \ldots $, let $\eta_{p|1} , \ldots , \eta_{p|p}$ denote the values of the
rvs $\eta_1, \ldots , \eta_{p}$ arranged in decreasing order, namely
$\eta_{p|p} \leq \ldots \leq \eta_{p|1}$,
with a lexicographic tiebreaker when needed. The rvs $\eta_{p|1} , \ldots , \eta_{p|p}$
are the {\em order statistics} associated with  the collection $\eta_{1} , \ldots , \eta_{p}$,
so that for each $s=1, \ldots , p$, the rv $\eta_{p|s}$ denotes the $s^{th}$ largest value amongst $\eta_{1} , \ldots , \eta_{p}$;
in particular $\eta_{p|1} $ and $\eta_{p|p}$ are the  maximum and minimum of the rvs 
$\eta_{1} , \ldots , \eta_{p}$, respectively \cite{DavidNagaraja, EKM}.

The evaluation of the limiting covariances (\ref{eq:LimitGivesC(d)}) proceeds with the following observation:
Fix $d=0,1, \ldots $ and take $n=3,4, \ldots $ such that $d \leq n-2$.
Under the enforced i.i.d. assumptions, for each $\theta > 0$ we get
\begin{eqnarray}
\lefteqn{
\left ( D^\star_{n,1}(\theta) , D^\star_{n,2}(\theta) \right )
} & &
\nonumber \\
&=&
\left (
\sum_{k=3}^n \1{ \xi_1 + \xi_k > \theta },  \sum_{k=3}^n \1{ \xi_2 + \xi_k > \theta }
\right )
\nonumber \\
&=_{st}&
\left (
\sum_{\ell=1}^{n-2} \1{ \xi_1 + \eta_\ell > \theta },  \sum_{\ell=1}^{n-2} \1{ \xi_2 + \eta_\ell > \theta }
\right )
\nonumber \\
&=&
\left (
\sum_{\ell=1}^{n-2} \1{ \xi_1 + \eta_{n-2|\ell} > \theta },  \sum_{\ell=1}^{n-2} \1{ \xi_2 + \eta_{n-2|\ell} > \theta }
\right )
\nonumber
\end{eqnarray}
where $=_{st}$ denotes distributional equality between rvs.
Two different cases arise:

First, with $d=0$ we find
\begin{eqnarray}
\bP{ D^\star _{n,1} (\theta) = 0  } 
&=&
\bP{ \sum_{\ell=1}^{n-2} \1{ \xi_1 + \eta_{n-2|\ell} > \theta } = 0  }
\nonumber \\
&=&
\bP{ \xi_1 + \eta_\ell \leq \theta,  \ \ell =1, \ldots , n-2  }
\nonumber \\
&=&
\bP{ \xi_1 +  \eta_{n-2|1} \leq \theta }
\nonumber \\
&=&
\bP{ \eta_{n-2|1}  - \theta \leq -\xi_1 }
\label{eq:CaseD=0ForOne}
\end{eqnarray}
and
\begin{eqnarray}
\lefteqn{\bP{ D^\star _{n,1} (\theta) = 0 , D^\star _{n,2} (\theta) = 0  } } & &
\nonumber \\
&=&
\bP{ 
\sum_{\ell=1}^{n-2} \1{ \xi_j + \eta_{n-2|\ell} > \theta } = 0, \ j=1,2
}
\nonumber \\
&=&
\bP{ \xi_1 + \eta_\ell \leq \theta ,  \xi_2 + \eta_\ell \leq \theta, \ \ell =1, \ldots , n-2 }
\nonumber \\
&=&
\bP{ \xi_1 +  \eta_{n-2|1} \leq \theta ,  \xi_2 +  \eta_{n-2|1} \leq \theta }
\nonumber \\
&=&
\bP{ \eta_{n-2|1} - \theta \leq  - \max \left ( \xi_1 , \xi_2 \right ) }.
\label{eq:CaseD=0ForTwo}
\end{eqnarray}

Next we consider the case $d=1,2, \ldots $.
Under the enforced independence assumptions we have
\begin{eqnarray}
\lefteqn{\bP{ D^\star _{n,1} (\theta) = d  } } & & 
\nonumber \\
&=&
\bP{ \sum_{\ell=1}^{n-2} \1{ \xi_1 + \eta_{n-2|\ell} > \theta } = d }
\nonumber \\
&=&
\bP{  \xi_1  +  \eta_{n-2|d+1} \leq \theta <  \xi_1 +  \eta_{n-2|d} }
\nonumber \\
&=&
\bP{ \theta -  \eta_{n-2|d} < \xi_1 \leq \theta -  \eta_{n-2|d+1}   }
\nonumber \\
&=&
\bE{
e^{- \lambda (  \theta -  \eta_{n-2|d}  )^+} - e^{- \lambda (  \theta -  \eta_{n-2|d+1}  )^+ }
} \hspace{10mm}
\label{eq:CaseDforOne}
\end{eqnarray}
and
 \begin{eqnarray}
\lefteqn{\bP{ D^\star _{n,1} (\theta) = d , D^\star _{n,2} (\theta) = d  } } & &
\nonumber \\
&=&
\bP{ 
\sum_{\ell=1}^{n-2} \1{ \xi_j + \eta_{n-2|\ell} > \theta } = 0, \ j=1,2
}
\nonumber \\
&=&
\bP{  \xi_j  +  \eta_{n-2|d+1} \leq \theta <  \xi_j  +  \eta_{n-2|d}, \ j=1,2 }
\nonumber \\
&=&
\bP{ \theta -  \eta_{n-2|d} < \xi_j \leq \theta -  \eta_{n-2|d+1} ,  \ j=1,2 }
\nonumber \\
&=&
\bE{
\left ( 
e^{- \lambda (  \theta -  \eta_{n-2|d}  )^+} - e^{- \lambda (  \theta -  \eta_{n-2|d+1}  )^+ }
\right )^2
}.
\label{eq:CaseDforTwo}
\end{eqnarray}

In the next step, carried out in Section \ref{sec:ProofPart+2}, 
we replace $\theta$ by $\theta^\star_n$ in the expressions above, and let $n$ go to infinity in the resulting expressions.
To evaluate these limits we shall rely on asymptotic properties of the order statistics 
which are discussed next.

\section{Asymptotic results for order statistics}
\label{sec:AsymptoticTheoryOrderStatistics}

We begin with a one-dimensional result.
For each $s=1, 2, \ldots $, consider the mapping
$G_s: \mathbb{R} \rightarrow \mathbb{R}_+$ defined by
\begin{equation}
G_s(x_s) 
\equiv \left ( \sum_{m=0}^{s-1} \frac{e^{-mx_s}}{m!} \right ) G(x_s),
\quad x_s \in \mathbb{R}
\label{eq:GeneralizedGumbel}
\end{equation}
where $G: \mathbb{R} \rightarrow \mathbb{R}_+$ denotes the well-known
{\em Gumbel} distribution given by
\begin{equation}
G(x) 
= e^{-e^{-x}},
\quad x \in \mathbb{R}.
\label{eq:GumbelCDF}
\end{equation}
The next result is well known \cite[Thm. 2.2.1, p. 33]{LeadbetterLindgrenRootzen}, and takes the following form when
applied to exponential distributions.

\begin{lemma}
{\sl For each $s=1,2, \ldots $, it holds that
\begin{equation}
\lim_{p \rightarrow \infty}
\bP{ \lambda \left ( \eta_{p|s} - \theta^\star_p \right ) \leq x_s }
= G_s (x_s),
\quad x_s \in \mathbb{R}
\label{eq:ConvergenceGeneralizedGumbelB}
\end{equation}
where the scaling $\theta^\star: \mathbb{N}_0 \rightarrow \mathbb{R}$ is given by (\ref{eq:FujiharaScaling}).
}
\label{lem:ConvergenceGeneralizedGumbel}
\end{lemma}

With $s=1$, Lemma \ref{lem:ConvergenceGeneralizedGumbel} expresses 
the well-known membership of exponential distributions in the maximal domain of attraction of the Gumbel distribution \cite{EKM}
\cite[Example 1.7.2, p. 21]{LeadbetterLindgrenRootzen}.

We now turn to the two-dimensional result we need:
For each $s=1, 2, \ldots $, define the mapping $J_s: \mathbb{R}^2 \rightarrow \mathbb{R}_+$ given by
\begin{eqnarray}
J_s(x_{s+1},x_s) 
\equiv 
\sum_{k_s=0}^{s-1} 
\left ( \ldots \right ) 
\times
\frac{ e^{ -k_s x_s } }{k_s!} \cdot
e^{- e^{- \min ( x_s , x_{s+1} )  } }
\label{eq:TwoDimLimitCDF+a}
\end{eqnarray}
with 
\begin{equation}
\ldots = \sum_{k_{s+1}=k_s}^{s}  
\frac{ \left ( e^{ -\min ( x_s , x_{s+1} ) } - e^{ -x_s } \right )^{k_{s+1} - k_s } }{ (k_{s+1} - k_s )!}
\label{eq:TwoDimLimitCDF+b}
\end{equation}
as $x_s$ and $x_{s+1}$ range over $\mathbb{R}$.
In these expressions we use the convention $0^0=1$.

\begin{lemma}
{\sl For each $s=1,2, \ldots$, we have
\begin{eqnarray}
& &\lim_{p \rightarrow \infty} 
\bP{  \lambda ( \eta_{p|s+1} - \theta^\star_p )  \leq x_{s+1}, \lambda ( \eta_{p|s} - \theta^\star_p )  \leq x_s }
\nonumber \\
&=&
J_s(x_{s+1}, x_s),
\quad x_{s+1}, x_{s} \in \mathbb{R}
\label{eq:KeyTwoDimAsymptotics}
\end{eqnarray}
with mapping $J_s: \mathbb{R}^2 \rightarrow \mathbb{R}_+$ given by (\ref{eq:TwoDimLimitCDF+a})-(\ref{eq:TwoDimLimitCDF+b}),
and scaling $\theta^\star: \mathbb{N}_0 \rightarrow \mathbb{R}$ given by (\ref{eq:FujiharaScaling}).
}
\label{lem:KeyTwoDimAsymptotics}
\end{lemma}

This result is a consequence of Theorem 2.3.1 in \cite[p. 34]{LeadbetterLindgrenRootzen}.
As only the case $s=1$ was discussed in \cite[Thm 2.3.2, p. 34]{LeadbetterLindgrenRootzen}, 
we provide in Section \ref{sec:ProofsAsymptoticsOrderStatistics} a proof for arbitrary 
values of $s$ when the variates $\{ \eta_k, \ k=1,2, \ldots \}$ are exponentially distributed.
By inspection we note from (\ref{eq:TwoDimLimitCDF+a})-(\ref{eq:TwoDimLimitCDF+b}) that
\begin{equation}
J_s(x_{s+1},x_s) = G_s( x_{s}) ,
\quad 
\begin{array}{c}
x_s \leq x_{s+1}  \\
x_s, x_{s+1} \in \mathbb{R} \\
\end{array}
\label{eq:TwoX_sBelowX_s+1}
\end{equation}
so that
\begin{align}
J_s(\infty ,x_s) &= \lim_{x_{s+1} \rightarrow \infty} J_s(x_{s+1},x_s) 
\nonumber \\
&= G_{s} (x_{s}),
\quad  x_{s} \in \mathbb{R} 
\label{eq:Marginal1}
\end{align}
while
\begin{align}
J_s(x_{s+1},\infty)
&=
\lim_{x_s \rightarrow \infty} J_s(x_{s+1},x_s) 
\nonumber \\
&= G_{s+1} (x_{s+1}),
\quad  x_{s+1} \in \mathbb{R} .
\label{eq:Marginal2}
\end{align}
This confirms that the probability distributions $G_{s+1}$ and $G_s$ are the one-dimensional marginal distributions of $J_s$ (as expected).

For use in Section \ref{sec:ProofPart+2} we find it convenient to give 
Lemma \ref{lem:ConvergenceGeneralizedGumbel} and Lemma \ref{lem:KeyTwoDimAsymptotics} 
the following probabilistic (and more compact)  formulation:
For any given  $s=1, 2, \ldots $,  there exists a pair  of $\mathbb{R}$-valued rvs $\Lambda_{s+1}$ and $\Lambda_{s}$ defined on 
$(\Omega , \mathcal{F}, \mathbb{P})$ such that
\begin{equation}
\lambda \left ( \eta_{p|s+j} - \theta^\star_p \right )  \Longrightarrow_p \Lambda_{s+j},
\quad j=0,1
\label{eq:oneDimConvergence}
\end{equation}
and
\begin{equation}
\left ( \lambda \left ( \eta_{p|s+1} - \theta^\star_p \right ),  \lambda \left ( \eta_{p|s} - \theta^\star_p \right ) \right )
\Longrightarrow_p ( \Lambda_{s+1}, \Lambda_{s})
\label{eq:JointConvergence}
\end{equation}
with $(\Lambda_{s+1}, \Lambda_{s})$ jointly distributed according to  $J_s$, 
and the $\mathbb{R}$-valued rvs $\Lambda_{s+1}$ and $\Lambda_{s} $
distributed according to $G_{s+1}$ and $G_{s}$, respectively.

\section{Completing the proof of Proposition \ref{prop:AssumptionC+fails}}
\label{sec:ProofPart+2}

We return to the expressions obtained in Section \ref{sec:ProofPart+1}:
With $d=0,1, \ldots $ held fixed, for each $n=2,3, \ldots$ we substitute $\theta$ by $\theta^\star_n$ in 
these expressions according to (\ref{eq:FujiharaScaling}),
and let $n$ go to infinity in the resulting expressions.

\subsection{The case $d=0$}
For each $n=3,4,\ldots $, with the aforementioned substitution,
we rewrite (\ref{eq:CaseD=0ForOne}) and (\ref{eq:CaseD=0ForTwo}) as
\[
\bP{ D^\star _{n,1} (\theta^\star_n) = 0  } 
=
\bP{ \lambda ( \eta_{n-2|1} - \theta^\star_{n} ) \leq - \lambda \xi_1 }
\]
and
\begin{eqnarray}
\lefteqn{ \bP{ D^\star _{n,1} (\theta^\star_n) = 0 , D^\star _{n,2} (\theta^\star_n) = 0  }  } & &
\nonumber \\
&=&
\bP{ \lambda ( \eta_{n-2|1} - \theta^\star_{n} )  \leq -  \lambda \max ( \xi_1, \xi_2 ) }
\nonumber
\end{eqnarray}
where by construction the rv $ \eta_{n-2|1} $ is independent of the i.i.d. rvs $\xi_1$ and $\xi_2$.

Let $\Lambda_1$ denote a rv  which is distributed according to the Gumbel distribution (\ref{eq:GumbelCDF}),
and which is independent of the i.i.d. rvs $\xi_1$ and $\xi_2$.
By Lemma \ref{lem:ConvergenceGeneralizedGumbel}  (for $s=1$ and $p=n-2$),
since $\lim_{n \rightarrow \infty} \left (  \theta^\star_n - \theta^\star_{n-2} \right ) = 0$, it is now plain  that
\begin{eqnarray}
p_{\rm Fuj}(0)
&=&
\lim_{n \rightarrow \infty}  \bP{ D^\star _{n,1} (\theta^\star_n) =  0 } 
\nonumber \\
&=& 
\bP{ \Lambda_1 \leq - \lambda \xi_1 }  = \bE{ e^{ -e^{  \lambda \xi_1  } } } 
\nonumber 
\end{eqnarray}
and
\begin{eqnarray}
\lefteqn{\lim_{n \rightarrow \infty} 
\bP{ D^\star _{n,1} (\theta^\star_n) = 0 , D^\star _{n,2} (\theta^\star_n) = 0  } }
& &\nonumber \\ 
&=&
\bP{ \Lambda_1 \leq - \lambda \max ( \xi_1, \xi_2) }  
= \bE{ e^{ -e^{  \lambda \max ( \xi_1, \xi_2 )  } } }
\nonumber 
\end{eqnarray}
under the independence assumptions.
Collecting these facts, we find
\begin{eqnarray}
\lefteqn{\lim_{n \rightarrow \infty}
{\rm Cov} \left [ \1{ D^\star_{n,1} (\theta^\star_n) = 0 } ,  \1{ D^\star_{n,2} (\theta^\star_n) = 0 } \right ] }
& &
\nonumber \\
&=&
\bE{ e^{ -e^{ \lambda \max ( \xi_1, \xi_2 )  } } } - \bE{ e^{ -e^{  \lambda \xi_1  } } } \bE{ e^{ -e^{  \lambda \xi_2  } } }.
\end{eqnarray}
As we make use of the reduction step (\ref{eq:LimitGivesC(d)}) discussed in Section \ref{sec:ProofPart+1}.
It follows that
\begin{eqnarray}
C(0)
&=&
\lim_{n \rightarrow \infty}
{\rm Cov} \left [ \1{ D^\star_{n,1} (\theta^\star_n) = 0 } ,  \1{ D^\star_{n,2} (\theta^\star_n) = 0 } \right ] 
\nonumber \\
&=&
\bE{ e^{ -e^{ \lambda \max ( \xi_1, \xi_2 )  } } } - \bE{ e^{ -e^{  \lambda \xi_1  } } } \bE{ e^{ -e^{  \lambda \xi_2  } } } 
\nonumber \\
&>& 0
\label{eq:C(0)}
\end{eqnarray}
since $\max( e^{\lambda \xi_1}, e^{ \lambda \xi_2} ) < e^{ \lambda  \xi_1  }  + e^{  \lambda \xi_2  } $.
\myendpf

\subsection{The case $d=1,2, \ldots $}

Pick $n=3,4,\ldots $ such that $d < n-2$.
Under the aforementioned substitution, we can rewrite
(\ref{eq:CaseDforOne}) and (\ref{eq:CaseDforTwo}) as
\begin{eqnarray}
\lefteqn{\bP{ D^\star _{n,1} (\theta^\star_n ) = d  } } & &
\nonumber \\
&=&
\bE{
e^{- \lambda (  \theta^\star_n  -  \eta_{n-2|d}  )^+} - e^{- \lambda (  \theta^\star_n -  \eta_{n-2|d+1}  )^+ }
}
\label{eq:CaseDforOneB}
\end{eqnarray}
and
 \begin{eqnarray}
 \lefteqn{ \bP{ D^\star _{n,1} (\theta^\star_n ) = d , D^\star _{n,2} (\theta^\star_n) = d  } } & &
\nonumber \\
&=&
\bE{
\left ( 
e^{- \lambda (  \theta^\star_n  -  \eta_{n-2|d}   )^+} - e^{- \lambda (  \theta^\star_n -  \eta_{n-2|d+1}  )^+ }
\right )^2
}.
\label{eq:CaseDforTwoB}
\end{eqnarray}

Applying Lemma \ref{lem:ConvergenceGeneralizedGumbel}  and Lemma \ref{lem:KeyTwoDimAsymptotics} (for $s=d$ and $p=n-2$) 
we conclude that

\[
\left ( \lambda \left ( \eta_{n-2|d+1} - \theta^\star_{n-2} \right ),  \lambda \left ( \eta_{n-2|d} - \theta^\star_{n-2} \right ) \right )
\Longrightarrow_n ( \Lambda_{d+1}, \Lambda_{d})
\]
in the notation used at (\ref{eq:JointConvergence}). 
Because $\lim_{n \rightarrow \infty} \left (  \theta^\star_n - \theta^\star_{n-2} \right ) = 0$,
we obtain
\begin{eqnarray}
\lefteqn{ \left ( \lambda \left ( \theta^\star_{n} - \eta_{n-2|d+1}  \right )^+,  \lambda \left ( \theta^\star_{n}  - \eta_{n-2|d} \right )^+  \right ) }
& & \nonumber \\ 
& &\Longrightarrow_n ( (-\Lambda_{d+1})^+ , (- \Lambda_{d})^+ ) \hspace{32mm}
%\label{eq:JointConvergence2}
\nonumber 
\end{eqnarray}
by the Continuous Mapping Theorem for weak convergence, whence
\begin{eqnarray}
\lefteqn{
e^{- \lambda (  \theta^\star_n  -  \eta_{n-2|d}  )^+} - e^{- \lambda (  \theta^\star_n -  \eta_{n-2|d+1}  )^+ } 
} & &
\nonumber \\
& & \Longrightarrow_n 
e^{-(-\Lambda_{d})^+ } - e^{-(-\Lambda_{d+1})^+ } 
\nonumber
\end{eqnarray}
by applying the Continuous Mapping Theorem once more.

Let  $n$ go to infinity in (\ref{eq:CaseDforOneB}) and (\ref{eq:CaseDforTwoB}): The Bounded Convergence Theorem yields
\begin{eqnarray}
\lefteqn{ \lim_{n \rightarrow \infty}
\bE{
\left (
e^{- \lambda (  \theta^\star_n  -  \eta_{n-2|d}  )^+} - e^{- \lambda (  \theta^\star_n -  \eta_{n-2|d+1}  )^+ }
\right )^a
} } & &
\nonumber \\
&=&
\bE{
\left (
e^{-(-\Lambda_{d})^+ } - e^{-(-\Lambda_{d+1} )^+ } 
\right )^a } \hspace{20mm}
\end{eqnarray}
for each $a=1,2$ upon observing the obvious bounds
\[
\left | e^{- \lambda (  \theta^\star_n  -  \eta_{n-2|d}  )^+} - e^{- \lambda (  \theta^\star_n -  \eta_{n-2|d+1}  )^+ } \right |
\leq 1,
\quad n=3,4, \ldots
\]
It follows that
\begin{eqnarray}
\lefteqn{ \lim_{n \rightarrow \infty}
{\rm Cov} \left [ \1{ D^\star _{n,1} (\theta^\star_n) = d}, \1{ D^\star _{n,2} (\theta^\star_n) = d  }  \right ]
} & &
\nonumber \\
&=&
\bE{
\left (
e^{-(-\Lambda_{d})^+ } - e^{-(-\Lambda_{d+1} )^+ } 
\right )^2
}
\nonumber \\
& &-
\left ( \bE{
e^{-(-\Lambda_{d})^+ } - e^{-(-\Lambda_{d+1} )^+ } 
}
\right )^2 \hspace{15mm}
\nonumber
\end{eqnarray}
and  the reduction step (\ref{eq:LimitGivesC(d)}) leads to
\begin{eqnarray}
C(d)
&=&
\lim_{n \rightarrow \infty}
{\rm Cov} \left [ \1{ D^\star _{n,1} (\theta^\star_n) = d}, \1{ D^\star _{n,2} (\theta^\star_n) = d  }  \right ]
\nonumber \\
&=&
{\rm Var}
\left [
e^{-(-\Lambda_{d})^+ } - e^{-(-\Lambda_{d+1})^+ } 
\right ] .
\label{eq:C(d)}
\end{eqnarray}
Note that $C(d) > 0$ as the variance of the non-degenerate rv
$e^{-(-\Lambda_{d})^+ } - e^{-(-\Lambda_{d+1})^+ } $.
\myendpf

\section{A Proof of Lemma \ref{lem:KeyTwoDimAsymptotics}}
\label{sec:ProofsAsymptoticsOrderStatistics} 

First some preliminaries.
Fix $p=1,2, \ldots $ and $u \geq 0$. The rv $S_p(u)$ given by
\[
S_p(u) = \sum_{\ell=1}^p \1{ \xi_\ell > u }
\]
counts the number of {\em exceedances} of level $u$ by the rvs $\xi_1, \ldots , \xi_p$.
The proof of  Lemma \ref{lem:KeyTwoDimAsymptotics} relies on the well-known equivalence 
\begin{equation}
\eta_{p|s} \leq u
\quad \mbox{if and only if} \quad
S_p(u) < s,
\quad s=0, 1, \ldots , p
\label{eq:equivalence1}
\end{equation}
given in \cite[Section 2.2, p. 33]{LeadbetterLindgrenRootzen}; see also \cite[Theorem 2.3.2, p. 36]{LeadbetterLindgrenRootzen}
for the case $s=1$.
Throughout we shall write 
\begin{equation}
u_p (x) = \lambda^{-1} \left ( \log p + x \right )^+ ,
\quad x \in \mathbb{R}.
\label{eq:u_p(x)}
\end{equation}

Fix $s=1, 2, \ldots $, and pick $x_s$ and $x_{s+1}$ in $\mathbb{R}$. Two cases are possible:

(i) If $x_s \leq x_{s+1}$ in $\mathbb{R}$, then for each $p=s+1,s+2, \ldots $,  it holds that $u_p(x_s)  \leq  u_p(x_{s+1}) $, whence
\begin{eqnarray}
\lefteqn{\bP{  \eta_{p|s+1} \leq u_p(x_{s+1}) , \eta_{p|s} \leq u_p(x_s)  } } & &
\nonumber \\
&=&
\bP{  \eta_{p|s+1} \leq u_p(x_s) , \eta_{p|s} \leq u_p(x_s)  } 
\nonumber \\
&=&
\bP{  \eta_{p|s} \leq u_p(x_s) }
\nonumber
\end{eqnarray}
since $\eta_{p |s+1} \leq \eta_{p|s} $.
From Lemma \ref{lem:ConvergenceGeneralizedGumbel} it follows that
\begin{eqnarray}
& &
\lim_{p \rightarrow \infty} 
\bP{  \lambda ( \eta_{p|s+1} - \theta^\star_p )  \leq x_{s+1}, \lambda ( \eta_{p|s} - \theta^\star_p )  \leq x_s } 
\nonumber \\
&=&
\lim_{p \rightarrow \infty} 
\bP{  \lambda ( \eta_{p|s} - \theta^\star_p )  \leq x_{s} }
\nonumber \\
&=&
G_{s} (x_s)
\nonumber
\end{eqnarray}
and (\ref{eq:KeyTwoDimAsymptotics}) holds as seen through (\ref{eq:TwoX_sBelowX_s+1}).

(ii) If $x_{s+1} \leq x_s$ in $\mathbb{R}$, then for each $p=s+1,s+2, \ldots $, 
the equivalence (\ref{eq:equivalence1}) yields
\begin{eqnarray}
\lefteqn{
\bP{  \eta_{p|s+1} \leq u_p(x_{s+1}) , \eta_{p|s} \leq u_p(x_s)  }
} & & 
\nonumber  \\
&=&
\bP{  S_p(u_p(x_{s+1})) < s+1,  S_p(u_p(x_s)) < s }
\label{eq:equivalence2} \\
\nonumber \\
&=&
\sum_{k_s=0}^{s-1} 
\left ( \sum_{k_{s+1}=0}^{s}  
\bP{  
\begin{array}{c}
S_p(u_p(x_{s+1})) = k_{s+1} \\
\\
S_p(u_p(x_s)) = k_s \\
\end{array}
}
\right )
 \label{eq:AA} \\
&=&
\sum_{k_s=0}^{s-1} 
\Bigg( \sum_{k_{s+1}=k_s}^{s}  
\bP{
\begin{array}{c}
 \sum_{\ell =1}^p \1{ \xi_\ell > u_p(x_{s+1}) }  = k_{s+1} \\
 \\
 \sum_{\ell =1}^p \1{ \xi_\ell > u_p(x_s) } = k_s 
\end{array}
}
\nonumber 
\end{eqnarray}
upon noting the fact $S_p(u_p(x_{s})) \leq S_p(u_p(x_{s+1}))$ since $u_p(x_{s+1})  \leq  u_p(x_{s}) $.
For arbitrary $k_s, k_{s+1} =0,1, \ldots $ with $k_s \leq k_{s+1}$, standard counting arguments give
\begin{eqnarray}
& &
\bP{ \begin{array}{c}
\sum_{\ell =1}^p \1{ \xi_\ell > u_p(x_{s+1}) }  = k_{s+1} \\
 \\
 \sum_{\ell =1}^p \1{ \xi_\ell > u_p(x_s) } = k_s  
 \end{array} 
 }
\nonumber \\
&=&
{ p \choose k_s } e^{- \lambda k_s u_p(x_s)}  \left ( e^{- \lambda u_p(x_{s+1})} - e^{- \lambda u_p(x_{s})} \right )^{k_{s+1} - k_s }
\nonumber \\
& & \times { p-k_s  \choose k_{s+1} - k_s } \left ( 1 - e^{- \lambda u_p(x_{s+1})} \right )^{p-k_{s+1}}
\nonumber \\
&=&
\frac{1}{k_s! ( k_{s+1} - k_s ) !} \cdot \frac{p! \cdot p^{- k_{s+1} } }{ (p-k_{s+1})!}
\cdot e^{ -k_s x_s } 
\nonumber \\
& & 
\times
\left ( e^{ -x_{s+1} } - e^{ -x_s } \right )^{k_{s+1} - k_s } 
\cdot
\left ( 1 - \frac{e^{-x_{s+1} } }{p} \right )^{p-k_{s+1}}
\label{eq:BB}
\end{eqnarray}
where the last step holds whenever $p$ is large enough so that $\log p + x_s > 0$ and $\log p + x_{s+1} > 0$.

Let $p$ go to infinity in (\ref{eq:AA}): From (\ref{eq:BB}) the limit of  each term in (\ref{eq:AA}) is given by
\begin{eqnarray}
\lefteqn{
\lim_{p \rightarrow \infty}
\bP{
\begin{array}{c}
\sum_{\ell =1}^p \1{ \xi_\ell > u_p(x_{s+1}) }  = k_{s+1} \\
 \\
\sum_{\ell =1}^p \1{ \xi_\ell > u_p(x_s) } = k_s  \\
\end{array}
}
} & &
\nonumber \\
%\label{eq:CCC} \\
&=&
\frac{ e^{ -k_s x_s } }{k_s!} 
\cdot
\frac{ \left ( e^{ -x_{s+1} } - e^{ -x_s } \right )^{k_{s+1} - k_s } }{ (k_{s+1} - k_s )!}
\cdot 
e^{- e^{-x_{s+1} } }
\nonumber
\end{eqnarray}
for $k_s \leq k_{s+1}$ in $\mathbb{N}$, 
%and $k_s , k_{s+1} = 0,1, \ldots$,
as we note that
\[
\lim_{p \rightarrow \infty}
\frac{p! }{ (p-k_{s+1})! \cdot p^{ k_{s+1} }}  =1 
\]
and 
\[
\lim_{p \rightarrow \infty} \left ( 1 - \frac{e^{-x_{s+1} } }{p} \right )^{p-k_{s+1}} = e^{- e^{-x_{s+1} } } .
\]
The convergence (\ref{eq:KeyTwoDimAsymptotics}) follows upon using the equivalence (\ref{eq:equivalence2})
together with the observation that $u_p (x_{s}) = \lambda^{-1} \left ( \log p + x_{s} \right )$ and
$u_p (x_{s+1}) = \lambda^{-1} \left ( \log p + x_{s+1} \right )$ for $p$ sufficiently large.
\myendpf

\section*{Acknowledgment}
The authors thank the anonymous referee from the first round of reviews for pointing out reference \cite{LeadbetterLindgrenRootzen} which lead to 
a much shorter proof of Proposition \ref{prop:AssumptionC+fails}, and for additional comments which greatly improved the presentation of the paper.
They also would like to thank another anonymous referee which indicated the possibility of extending the original results 
to a more general setting without homogeneity assumptions as was done in Section \ref{sec:GeneralSettingLittleTheory}.

\begin{IEEEbiography}[{\includegraphics[width=1in,height=1.25in,clip,keepaspectratio]{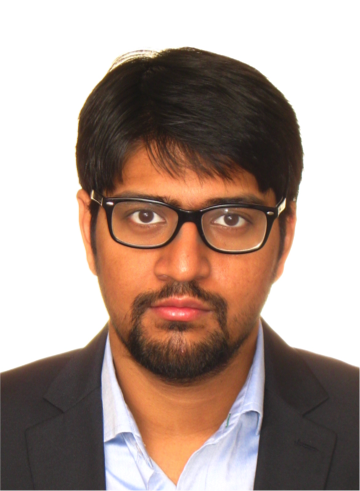}}]{Siddharth Pal}
received his Bachelors degree in Electronics and
Telecommunication Engineering from Jadavpur University, India, in
2011, and his Masters and Ph.D degrees. both in Electrical Engineering
from University of Maryland College Park, USA, in 2014 and 2015 respectively. Since then he has been working as a research scientist
at Raytheon BBN Technologies. His research interests include network
science and analysis, game theory, machine learning with emphasis on
neural network based approaches, and stochastic systems.
\end{IEEEbiography}

\begin{IEEEbiography}[{\includegraphics[width=1in,height=1.25in,clip,keepaspectratio]{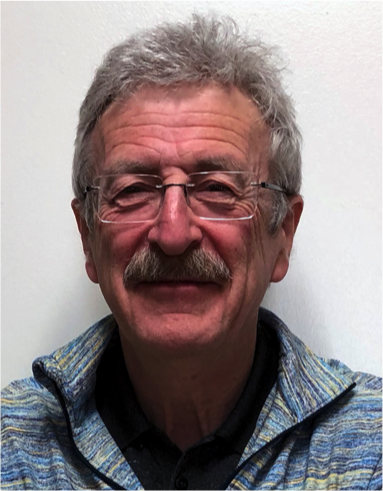}}]{Armand M. Makowski} 
(M'83√±-SM'94√±-F'06)
received the Licence en Sciences Math\'ematiques from the
Universit\'e Libre de Bruxelles in 1975, the M.S. degree in
Engineering-Systems Science from U.C.L.A. in 1976 and the Ph.D.
degree in Applied Mathematics from the University of Kentucky in
1981. In August 1981, he joined the faculty of the Electrical
Engineering Department at the University of Maryland College Park,
where he is Professor of Electrical and Computer Engineering. He
has held a joint appointment with the Institute for Systems
Research since its establishment in 1985.

Armand Makowski was a C.R.B. Fellow of the Belgian-American
Educational Foundation (BAEF) for the academic year 1975-76; he is
also a 1984 recipient of the NSF Presidential Young Investigator
Award and became an IEEE Fellow in 2006.

His research interests lie in applying advanced methods from the
theory of stochastic processes to the modeling, design and
performance evaluation of engineering systems, with particular
emphasis on communication systems and networks.
\end{IEEEbiography}

\end{document}